\documentclass[amssymb, reprint, showpacs, aip, groupedaddress]{revtex4-1}	

\usepackage{graphicx}				
\usepackage{dcolumn}
\usepackage{bm}
\usepackage[colorlinks=true,linkcolor=blue,citecolor=blue]{hyperref}
\usepackage{float}
\usepackage{amsmath, amsfonts}
\usepackage{url}
\usepackage{setspace}
\usepackage{lineno}
\usepackage{gensymb}
\usepackage{subfigure}
\usepackage{multirow}
\usepackage{array}
\usepackage[normalem]{ulem}
\usepackage[toc,page]{appendix}
\usepackage{chngcntr}
\usepackage{bbding}

\graphicspath{{./figures/}}

\date{\today}							

\begin{document}

\preprint{JASA/123}

\title[Eastern Arctic ambient noise]{Eastern arctic ambient noise on a drifting vertical array}

\author{Emma Reeves}
\email[Corresponding author. Email: ]{ecreeves@ucsd.edu}
\author{Peter Gerstoft}
\author{Peter F. Worcester}
\author{Matthew A. Dzieciuch}
\author{Aaron Thode}

\affiliation{Scripps Institution of Oceanography, University of California San Diego, La Jolla, California 92093--0238}

\date{\today}

\begin{abstract}
Ambient noise in the eastern Arctic was studied from April to September 2013 using a 22 element vertical hydrophone array as it drifted from near the North Pole (89\degree 23'N, 62\degree 35'W) to north of Fram Strait (83\degree 45'N  4\degree 28' W). The hydrophones recorded for 108 min/day on six days per week with a sampling rate of 1953.125 Hz. After removal of data corrupted by non--acoustic transients, 19 days throughout the transit period were analyzed. Noise contributors identified include broadband and tonal ice noises, bowhead whale calling, seismic airgun surveys, and earthquake $T$ phases. The bowhead whale or whales detected are believed to belong to the endangered Spitsbergen population and were recorded when the array was as far north  as 86$\degree$24'N. Median power spectral estimates and empirical probability density functions (PDFs) along the array transit show a change in the ambient noise levels corresponding to seismic survey airgun occurrence and received level at low frequencies and transient ice noises at high frequencies. Median power for the same periods across the array show that this change is consistent in depth. The median ambient noise for May 2013 was among the lowest of the sparse reported observations in the eastern Arctic but comparable to the more numerous observations of western Arctic noise levels.

\end{abstract}


\maketitle


\section{\label{sec: intro} Introduction}

Ambient noise in the Arctic Ocean is strongly influenced by its sea ice cover and upward refracting sound speed profile. Internal frictional shearing, thermal stress fracturing, and interaction within leads in the ice generate distinct sounds that are received acoustically at levels exceeding 100 dB re 1$\mu$ Pa$^2$ Hz$^{-1}$. The widespread ice cover deters many animal species from venturing far north, but attracts species capable of seeking ice leads or generating their own breathing holes, such as bowhead whales.\cite{MooreReeves1993} At the same time, the upward--refracting sound speed profile and nearly year--round ice cover allow low frequency signals to propagate long distances while attenuating higher frequency components. This unique environment depends strongly on the properties of the Arctic sea ice, including percentage of areal cover, thickness (age), under--ice roughness, and lateral extent. Over the past decade, the Arctic sea ice has dramatically reduced in thickness as well as annual extent,\cite{LindsaySchweiger} resulting in unknown changes to the ambient noise environment that this study investigates through use of recent data and analysis.

Sea ice noise and the Arctic ambient noise properties have historically been an area of interest in underwater acoustics.\cite{Dyer84}$^,$\cite{Cummings89} Measurements of transient ice noises have shown that they are highly non--Gaussian,\cite{VeitchWilks85} varying in frequency, bandwidth, length, and received sound level according to the sea ice properties and environmental conditions,\cite{Kinda2015} but are often more prevalent near ice ridges.\cite{Diachok76}$^,$\cite{BuckWilson86} The cumulative ambient noise levels generated by ice noise have been shown to correlate with environmental variables like wind, air pressure, and temperature.\cite{GreeneBuck64}$^,$\cite{MakrisDyer86}$^,$\cite{MakrisDyer91}$^,$\cite{Kinda2013} Near the Marginal Ice Zone (MIZ), where the ice is subject to increased wave forcing, noise levels have been shown to be as much as 10 dB higher than those further away from the MIZ.\cite{Diachok1974}$^,$\cite{Geyer} Sea ice is a strong scatterer that attenuates high frequencies at a much higher rate than the open ocean,\cite{Diachok80} although the exact attenuation coefficients depend on the local sea ice structure in ways that have yet to be determined.\cite{TollefsenSagen} Due in large part to biological activity and experimental accessibility, the Western Arctic ambient noise near the Beaufort Sea\cite{Kinda2013}$^,$\cite{Roth}$^,$\cite{NOAA_Arctic}$^,$\cite{LewisDenner87seasonal} has been studied more extensively than the eastern Arctic ambient noise (defined here as areas east of 60$\degree$W). Studies north of 85$\degree$N are extremely rare.\cite{MooreReeves1993}

In April 2013, a bottom--moored vertical hydrophone array was deployed at Ice Camp Barneo near 89\degree N, 62\degree W. The experiment was designed to study acoustic propagation and ambient noise under the sea ice. Around April 15 the mooring cable failed. The subsurface float rose to the surface and remained there, with the array hanging unweighted below. It drifted southward with the Transpolar Current toward the Fram Strait, recording ambient noise as scheduled. MicroCAT pressure measurements (see Sec.~\ref{subsec: microcats}) showed that the array was vertical under its own weight during much of the transit. The resulting data record the spatiotemporal variation of the far northern Arctic ambient noise ($>$ 85 \degree N). In this study, the dataset is analyzed and the observations are interpreted in terms of previous studies of this ambient noise.

This paper is organized as follows. In Sec.~\ref{sec: data}, the acoustic experiment is described, data processing methods are explained, and the collection of supplementary environmental data is discussed. Sec.~\ref{sec: contrib} discusses select noise events. Sec.~\ref{sec: results} presents the results of statistical ambient noise analyses in both time and depth, and Arctic ambient noise power estimates from previous studies are compared with the results. The goal of this paper is to establish an understanding of ambient noise contributors and sound levels in the northeastern Arctic during summer 2013.

\section{\label{sec: data} Methods}
\subsection{\label{subsec: acoustics} Acoustic measurements}

      A 600 m long bottom-moored acoustic receiving array was deployed at Ice Camp Barneo, 89\degree 23'N, 62\degree 35'W, on April 14. Twenty-two omnidirectional hydrophone modules (H.M.) were spaced along the array, with H.M. 1--10 separated by 14.5 m and H.M. 11--22 separated by logarithmically increasing spacing starting at 16.5 m (Table \ref{table: array}, Fig.~\ref{fig: Map_Float}). The topmost hydrophone was 11.6 m below the subsurface float. The hydrophones recorded underwater sound for 108 min/day six days per week, starting at 1200 UTC each day, with a sampling frequency of 1,953.125 Hz. The hydrophone recording schedule was constrained by the amount of data storage available in the hydrophone modules. Acoustic recordings are available for 119 days between April 29 and September 20.
      
                     \begin{table}[h]
             \begin{center}
             \caption{\label{table: array} Instrument spacing, numbering, and MicroCAT sampling periods for the instruments on the VLA during its drifting period. The depth estimates assume that the subsurface buoy was floating at 0 m, an assumption confirmed by the MicroCAT measured depths.}
             \begin{tabular}{ p{1.4cm}<{\centering} p{1.7cm}<{\centering} p{1.7cm}<{\centering} p{1.7cm}<{\centering} p{1.7cm} <{\centering}}
             \hline\hline
           H.M. \# &  H.M. Depth (m) & MicroCAT Depth (m) & MicroCAT \# & MicroCAT Sampling Period (s)\\ 
             \hline
	    1 & 11.6 & 4.6 & 1 & 480 \\ 
             2 & 26.1 & 24.6 & 2 & 480\\ 
             3 & 40.6 &  &  & \\ 
             4 &  55.1 & 49.6 & 3 & 480 \\ 
             5 & 69.6 &  &  & \\ 
             6 & 84.1 &  &  & \\ 
             7 & 98.6 & 99.6 & 4 & 480\\ 
             8 & 113.1 &  &   & \\ 
             9 & 127.6 &  &  & \\ 
             10 & 142.1 &  &  & \\ 
             11 & 158.7 & 149.6 & 5 & 380\\ 
             12 & 177.7 &  &  & \\ 
             13 & 199.5 & 200.6 & 6 & 380 \\ 
             14 & 224.4 &  &  &  \\ 
             15 & 253 &  249.6 & 7 & 380 \\ 
             16 & 285.7 &  &  &  \\ 
             17 & 323.2 &  &  & \\
             18 & 366.1 & 349.6 & 8 & 380 \\ 
             19 & 415.2 &  &  & \\ 
             20 & 471.4 & 449.6 & 9 & 380 \\
             21 & 535.8 &  & &  \\ 
             22 & 609.6 & 599.6 & 10 & 300\\ 
             \hline\hline
             \end{tabular}
             \end{center}
             \end{table}
      
      The raw acoustic recordings were scaled to be in units of instantaneous sound pressure using the analog-to-digital conversion parameters, the gain, and the hydrophone receiving sensitivity given by the manufacturer. The hydrophone receiving sensitivity was nearly constant above 50 Hz but highly frequency dependent below 50 Hz. The system noise floor was computed using a model that combines the known self--noise of its individual components. The system was experimentally tested in a Faraday cage and by calculating the coherence between multiple sensors recording noise in a quiet room. Both tests fit the modeled system noise floor well.

Median (50\%) spectral estimates were created by segmenting three or four day periods of data (see Sec.~\ref{subsec: NoiseProcessing}) into 4096-point windows  ($\sim$ 2 s), taking a 16,384-point Fast Fourier Transform to interpolate to high resolution frequency bins of 0.12 Hz, and sorting the individual spectral estimates by power level at each frequency bin. The probability density (PDF) was estimated from these spectral estimates using 100 power bins of equal width at each frequency.  The PDF for a target frequency was obtained by averaging three PDFs closest to the target frequency. Spectrograms were estimated using shorter, 512-point windowed segments ($\sim$ 0.25 s) zero-padded to 2048 points (df $\approx$ 1 Hz) in order to capture transients of length $<1$ s. Unless otherwise noted, the data were recorded at 84.1 m depth (hydrophone \# 6) for comparability to other ambient noise studies in the eastern Arctic.\cite{MakrisDyer86}

\subsection{\label{subsec: microcats} MicroCATs}
       Ten Sea-Bird SBE 37--SM/SMP MicroCAT instruments, measuring temperature, conductivity, and pressure (dBars) were co-located with the hydrophones, spaced 25, 50, 50, 50,  50, 100, 100, and 150 m apart. The topmost MicroCAT was located 4.6 m below the subsurface float (Table \ref{table: array}, Fig.~\ref{fig: Map_Float}). The MicroCATs began recording on April 28 and sampled continuously until September 19. The sampling period for each MicroCAT is shown in Table \ref{table: array}.

\subsection{GPS coordinates}

\begin{figure*}[ht]
\subfigure{
\includegraphics[width =  0.7\linewidth]{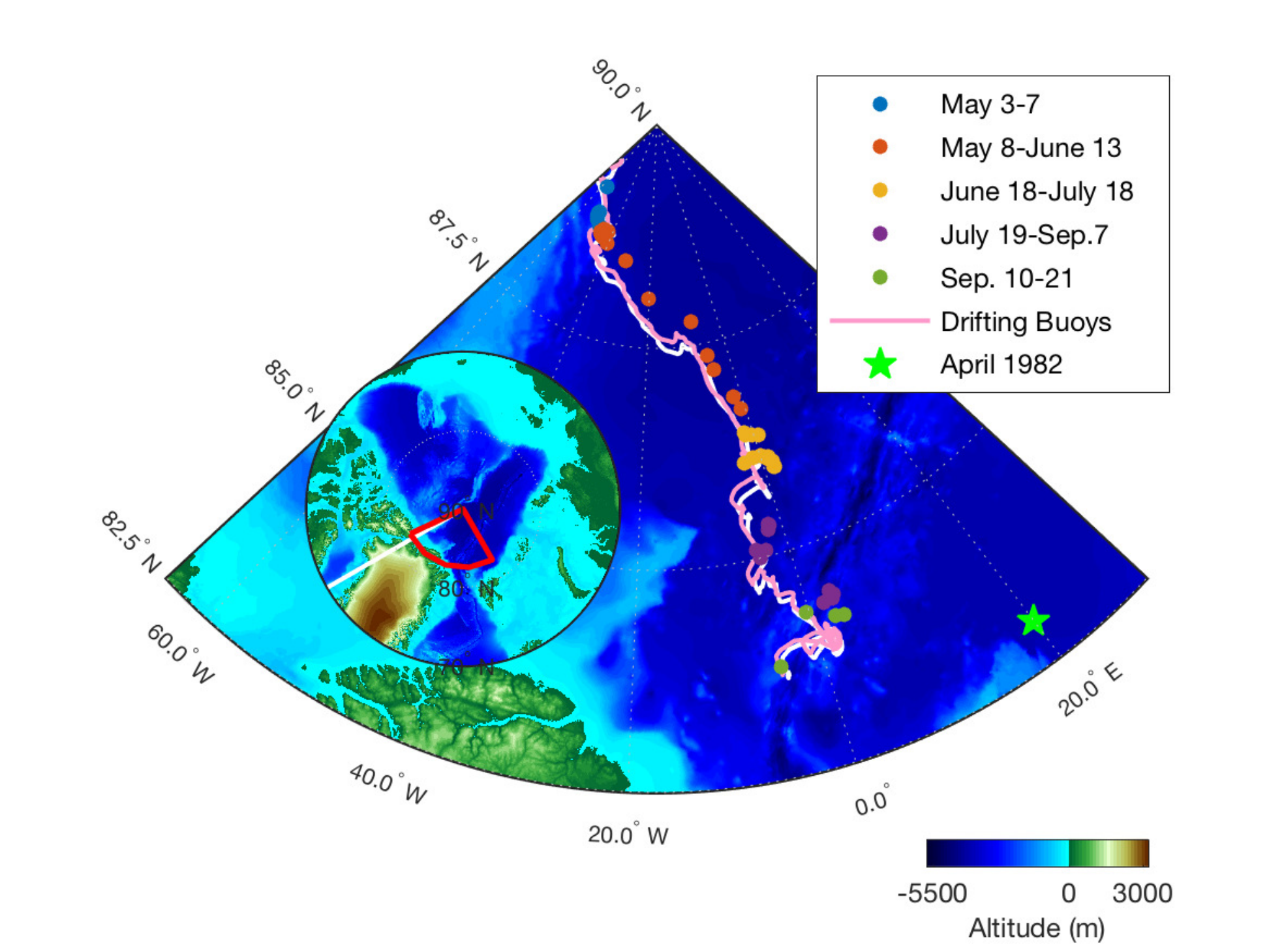}\label{fig: map}}
\subfigure{
\includegraphics[width = 0.28\linewidth]{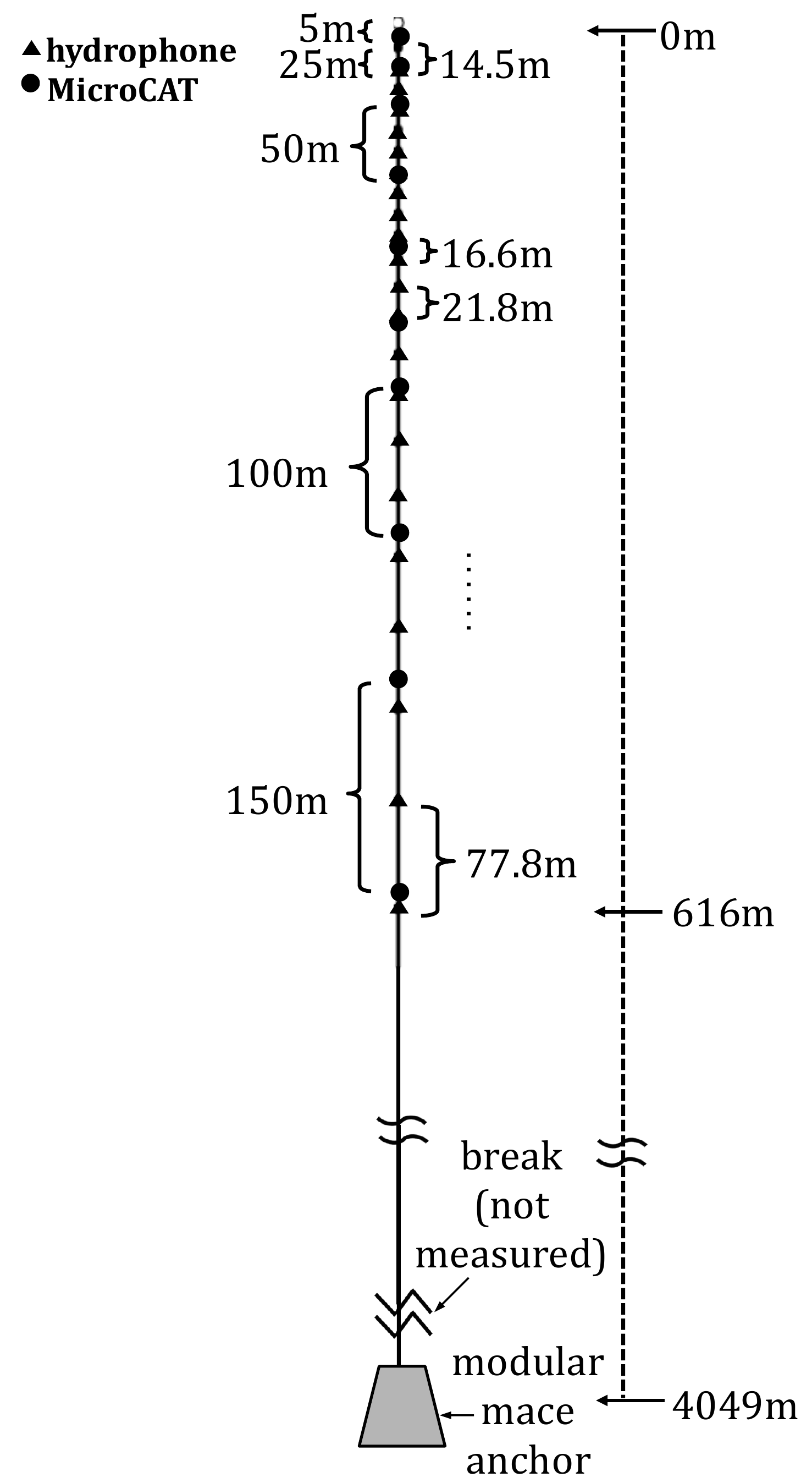}\label{fig: array}}
\caption{(Color online) A bathymetric relief map displaying the location of the receiving array divided according to hydrophone processing period (see Sec.~\ref{sec: results}) along with the location of two concurrently deployed ice--moored buoys with daily GPS and the April 1982 FRAM IV ice camp (\FiveStar). A map inset shows the location of the array path relative to the Arctic and a line indicating the 60\degree W longitude.  The moored array design is shown to the right of the map.}
\label{fig: Map_Float}
\end{figure*}

 A Xeos Technologies Kilo Iridium-GPS mooring location beacon located on top of the subsurface float began transmitting ALARM messages on May 3, indicating that the mooring had prematurely surfaced. The reported position at the time of surfacing was 88\degree 50'N, 51\degree 17'W, 63 km from the deployment location. Analysis of an acoustic survey on April 14, following deployment of the mooring, revealed that the acoustic release was significantly shallower than expected. The implication is that the mooring failed shortly after deployment, but the subsurface float was trapped beneath sea ice, preventing the location beacon from obtaining  GPS positions or transmitting ALARM messages until it was exposed on May 3.  The float drifted southward in the Transpolar Drift. There were frequent gaps in transmissions from the location beacon which are presumed to coincide with periods when the subsurface float was covered by sea ice. The buoy was recovered on September 21, at 84\degree 03'N, 03\degree 05'W. The mooring line was found to have parted immediately above the anchor (Fig.~\ref{fig: Map_Float}). 

     \subsection{Bathymetry}
     
  	The International Bathymetric Chart of the Arctic Ocean from the National Centers for Environmental Information was used to construct a map of the ocean depth relative to the array location (Fig.~\ref{fig: Map_Float}).
	
	The measured depths varied between 2.5 km and 4.7 km during the drift period. The Gakkel Ridge was the shallowest area crossed by the array, and it is possible that the array interacted with the bottom there or in other shallow regions. Without instrumentation on the lower array, the presence of array--bottom interaction cannot be determined.

\subsection{Sea Ice Concentration}\label{sec: ice}
Daily sea ice concentration, defined as the areal percentage of satellite imagery above a certain brightness level, was obtained from the Advanced Microwave Scanning Radiometer-2 (AMSR-2) 89-GHz channel satellite dataset,\cite{AMSR2} provided in a 4 km X 4 km gridded format from the Institute of Environmental Physics, University of Bremen, Germany. The sea ice concentration ranges from 0 (no ice) to 100 (solid ice). The georeferenced latitude and longitude grids were transformed into regular latitude and longitude grids with 0.1$\degree$ resolution with the ice concentration interpolated to the array location.

In addition, the AMSR-2 satellite data were used to determine the daily distance from the array to the ice edge. This distance was about 1000 km in April and 200 km in September, decreasing steadily as the array drifted closer to the MIZ.

\subsection{Filtering/Noise Removal}\label{sec: filtering}

 \begin{figure}[]
\centering
\subfigure{
\includegraphics[width = \linewidth]{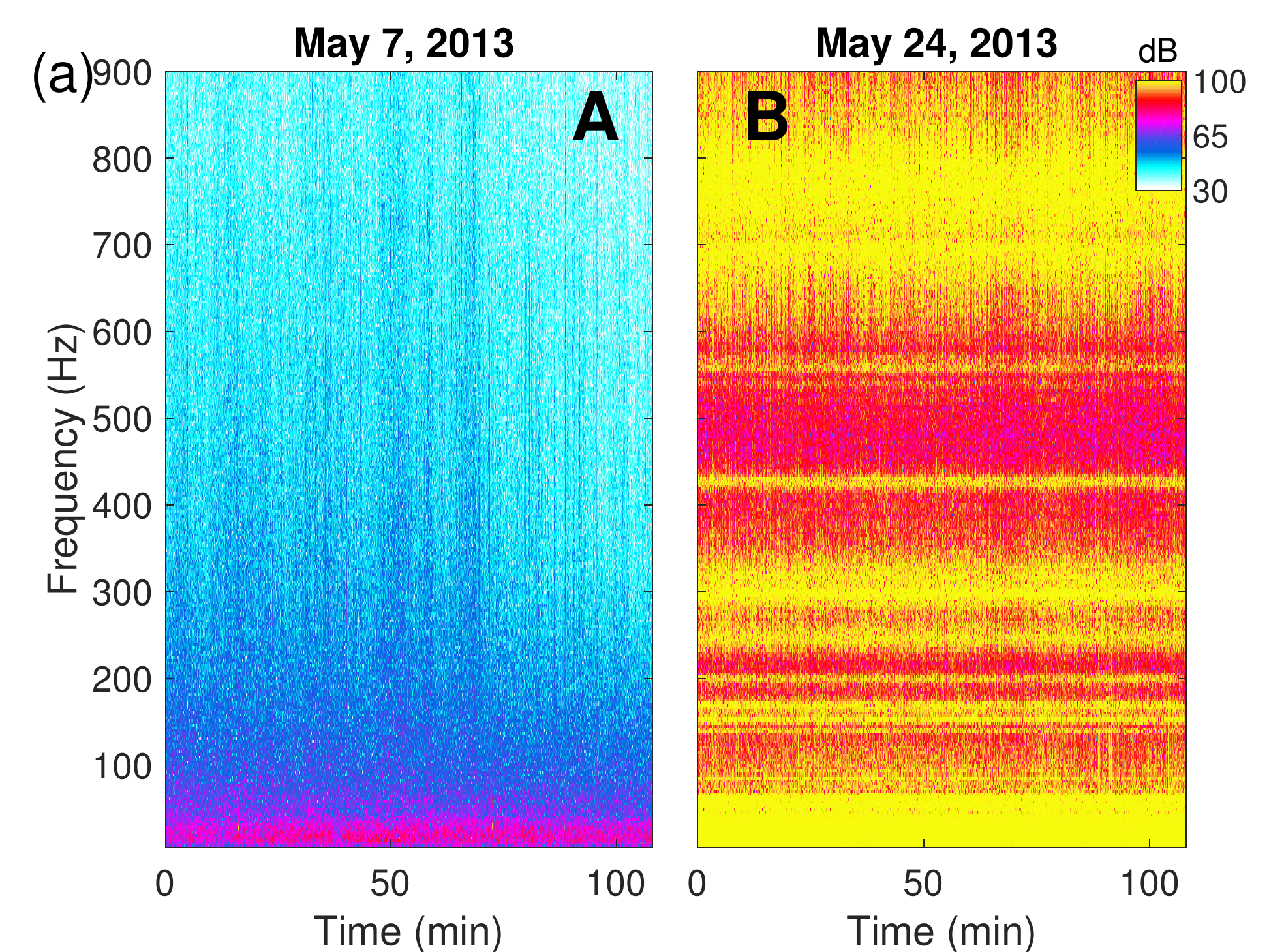}\label{fig: noise_artifacts}}
\subfigure{
\includegraphics[width = \linewidth]{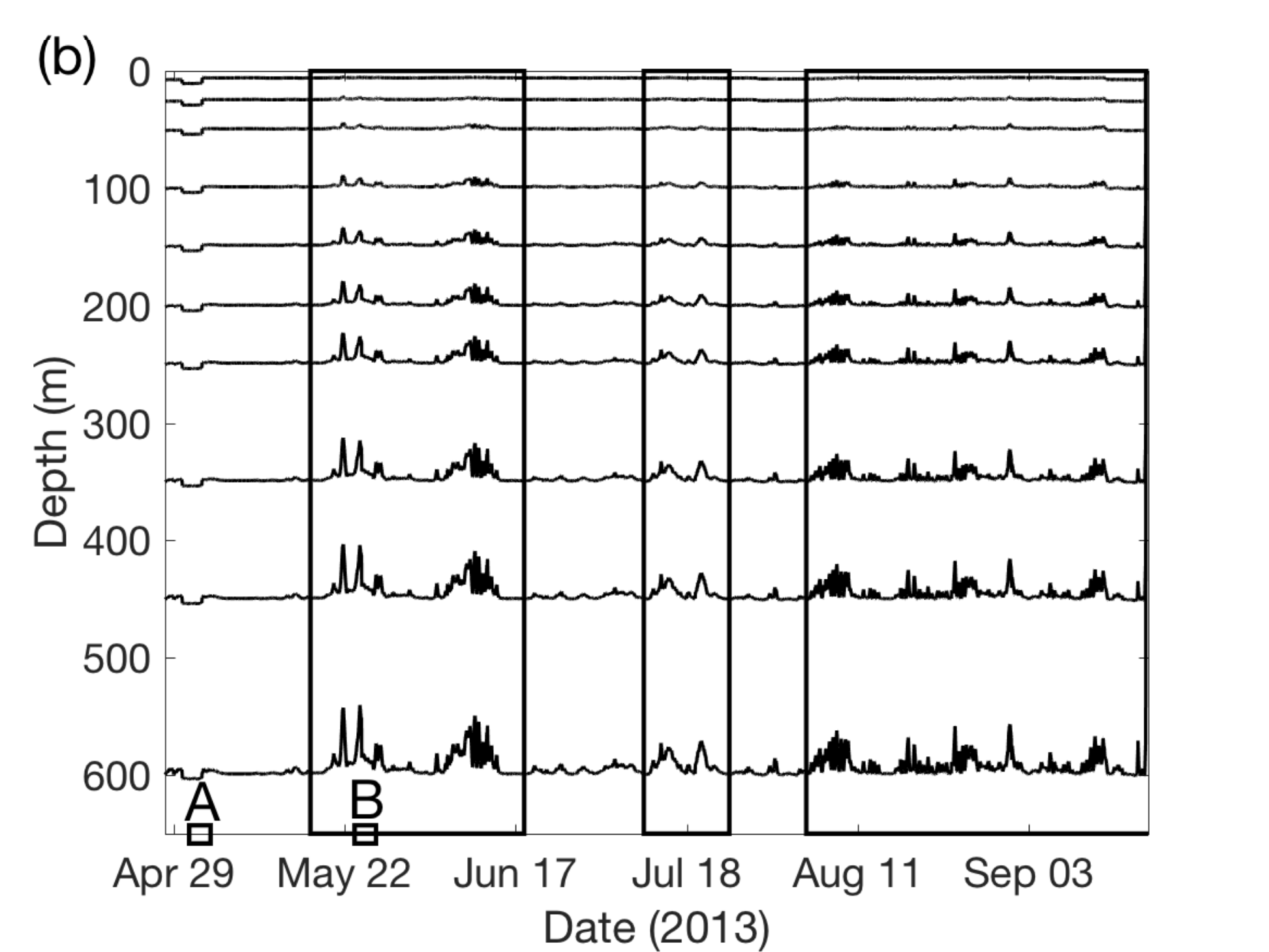}\label{fig: pressures}}
\caption{(Color online) (a) Spectrograms generated from hydrophone recordings at 609.6 m depth during periods containing typical ambient noise (\textbf{A}) and strong spectral bands considered non--acoustic artifacts (\textbf{B}). (b) MicroCAT pressure measurements at ten depths exhibit periods of shallowing (rectangles) that correspond to artifacts in (a).}\label{fig: filters}
\end{figure}

	The drifting array was heavily contaminated by self--noise at certain times. Low frequency ($f<5$ Hz) cable strum was observed. Strong spectral bands were also observed, exceeding 100 dB re 1 $\mu$Pa$^2$ Hz$^{-1}$ and extending to the Nyquist frequency (976.56 Hz). These elevated spectral levels, predominant in the frequency bands 0--50 Hz, 250--325 Hz, and 600--900 Hz (Fig.~\ref{fig: noise_artifacts}), were found to correspond with periods of unexpectedly low pressures (depths) on the MicroCATs (Fig.~\ref{fig: pressures}), making them unlikely to be caused by propagating acoustic noise and more likely to be noise artifacts. With the buoyant subsurface float constrained to the surface, flow past the mooring lifts and thus tilts the array and reduces the MicroCAT pressures (depths). Potential non-acoustic noise sources on the mooring, which lacked fairing, include strumming--induced vibration, flow noise, and/or bottom interaction. The noise artifacts were not correlated with bathymetry or wind reanalysis data and could not be removed by a $\omega$--k beamforming filter indicating that the instruments were directly affected.

To remove affected data, the median MicroCAT pressure for each day was computed. The pressure on MicroCAT \#10 (599.6 m) had the largest variation between days and was used as an indicator of flow-related noise. By comparing the good and bad  spectrograms with the median pressures on MicroCAT \#10  (Fig.~\ref{fig: filters}) , it was found that most corrupted data had a median MicroCAT pressure of less than 604.9 dBars. Therefore days with $p_{\text{MicroCAT,10}} < 604.9$ dBars were not used. This method selected 19 days for further analysis: April 30, May 1, 2, 7, 8, 9, 12, 14, June 16, 18, July 3, 14, 19, 24, August 2, and September 10, 18, 19, 20. There is evidence that the noise artifacts were not completely removed for one or two periods (see Sec.~\ref{sec: results}).

\section{Arctic ambient noise source effects}\label{sec: contrib}
\subsection{Underwater Sound Propagation}\label{sec: prop}

Eastern Arctic ambient noise is influenced by the characteristics of sound propagation which are affected by the oceanographic water masses and sea ice cover in the region.\cite{Mikhalevsky2001} Much of this propagation is over long distances due to the intermittent nature of nearby ice noise events (see Sec.~\ref{sec: Icenoise}), the infrequency of biological activity (see Sec.~\ref{sec: Bio}), and the locations of regular anthropogenic activity (see Sec.~\ref{sec: Airguns}).

\begin{figure}[]
     \centering
     \subfigure{
	\includegraphics[width = \linewidth]{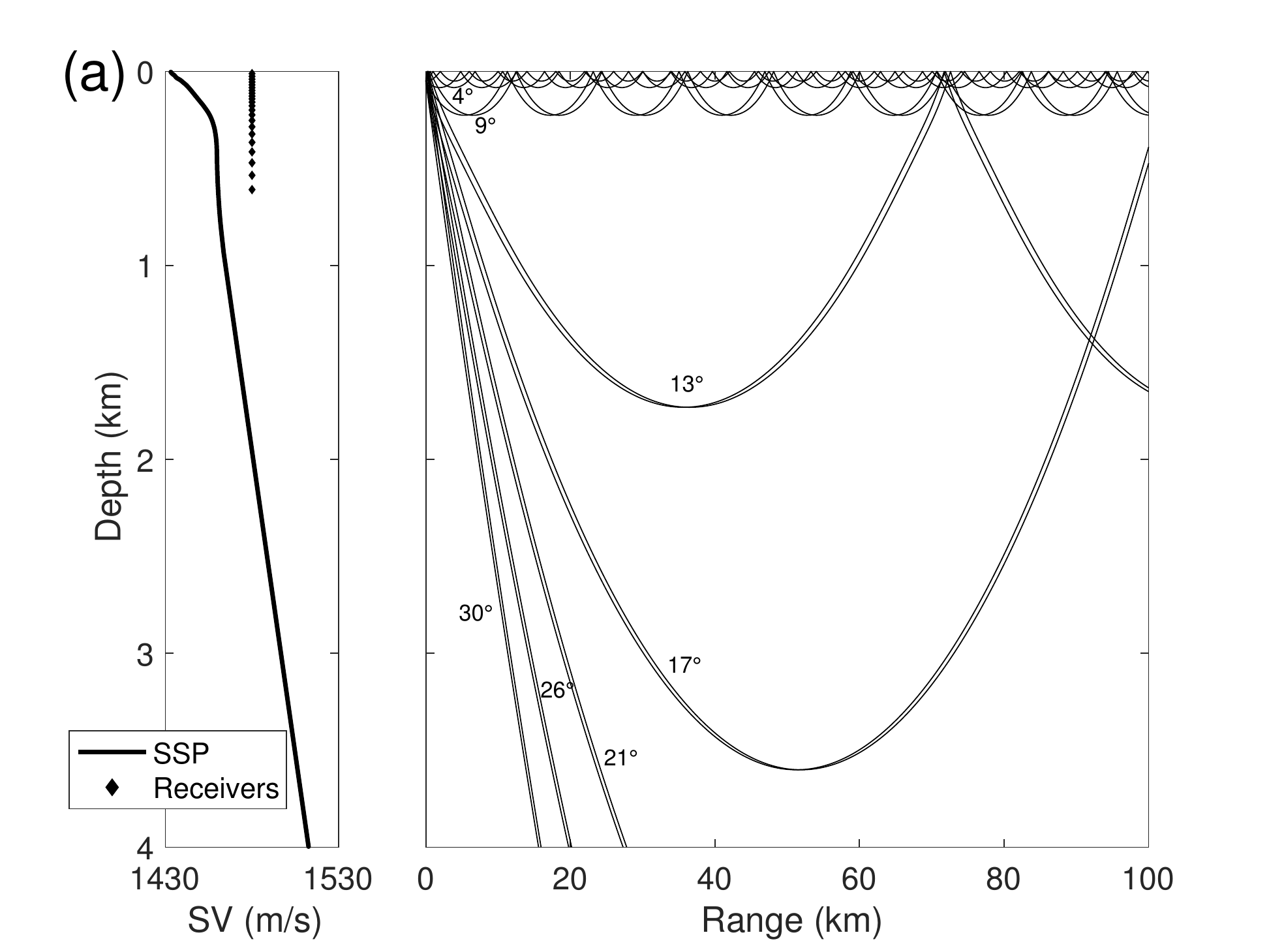}\label{fig: SSP}}
	\subfigure{
\includegraphics[width =  \linewidth]{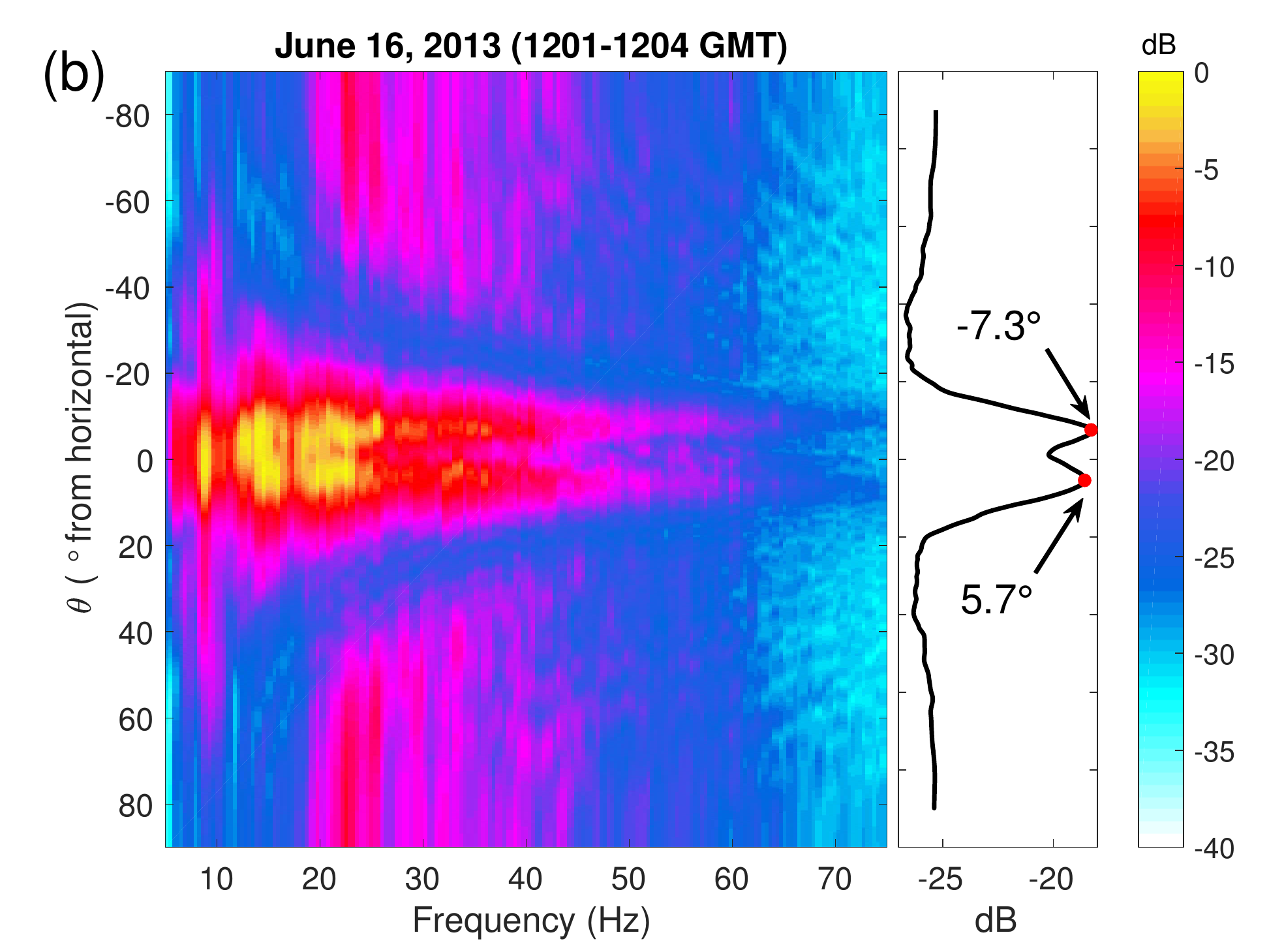}\label{fig: beam}}
     \caption{ (a) Bellhop ray propagation model \cite{Jensen} for a near--surface source using the sound speed profile measured at Ice Camp Barneo demonstrates the strongly upward refracting profile. Rays were launched between $\pm$30\degree  from horizontal. (b) Bartlett beamformer at received airgun pulse frequencies, averaged across 1201--1204 GMT on June 16. The arrivals at $-7\degree$ and 5$\degree$ indicate the preservation of intermediate ray angles over long range propagation ($\theta< 0\degree$ is upward--looking).}
     \end{figure}

The sound speed profile in the eastern Arctic is strongly upward refracting with a minimum at the ocean--ice interface (Fig.~\ref{fig: SSP}). The relevant water masses include Polar Water (0--200m), Arctic Intermediate Water (AIW, 200--1000m), and Deep Polar Water ($>$1000m).\cite{Mikhalevsky2001} Profiles in the eastern Arctic differ from western Arctic in that the depth of the AIW temperature maximum is considerably shallower in the eastern Arctic.

In completely ice--covered environments, the sea ice acts as a low--pass filter.\cite{Mikhalevsky2001} Higher frequency sound ($f > 30$ Hz and $\lambda < 50$ m) is strongly scattered at the water--ice interface. In addition, the number of reflections from the sea ice per kilometer increases as a propagating ray's angle decreases ($<$5\degree, Fig.~\ref{fig: beam}).

\begin{figure}[]
\centering
\includegraphics[width = \linewidth]{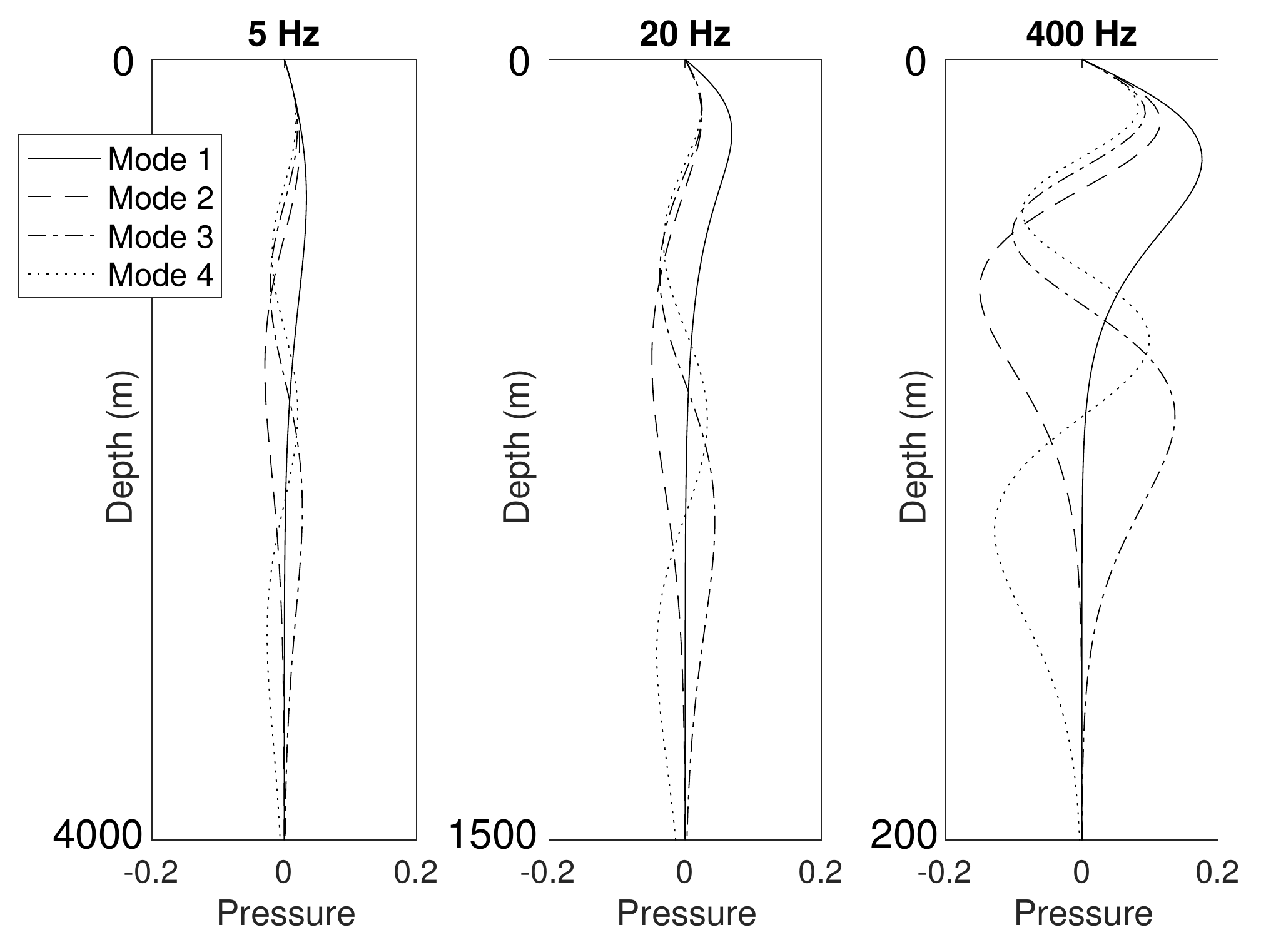}
\caption{Modal structure of the eastern Arctic environment at three frequencies. Each panel has a depth scale appropriate for the vertical scale of the modes at that frequency.}\label{fig: Modes}
\end{figure}

On the other hand, steeper rays ($>$ about 13--15\degree) experience fewer reflections per kilometer but will interact with bathymetric features, especially at the Gakkel Ridge where the ocean depth shallows to nearly 2 km (Fig.~\ref{fig: map}). At low frequencies (Fig.~\ref{fig: Modes} at 5 Hz), even the lowest modes interact with and scatter from bathymetric features, leading to lower ambient noise levels below 10 Hz.

\subsection{Ice--generated noise\label{sec: Icenoise}}

\begin{figure*}[ht]
\centering
\subfigure{
\includegraphics[width = 0.48\textwidth]{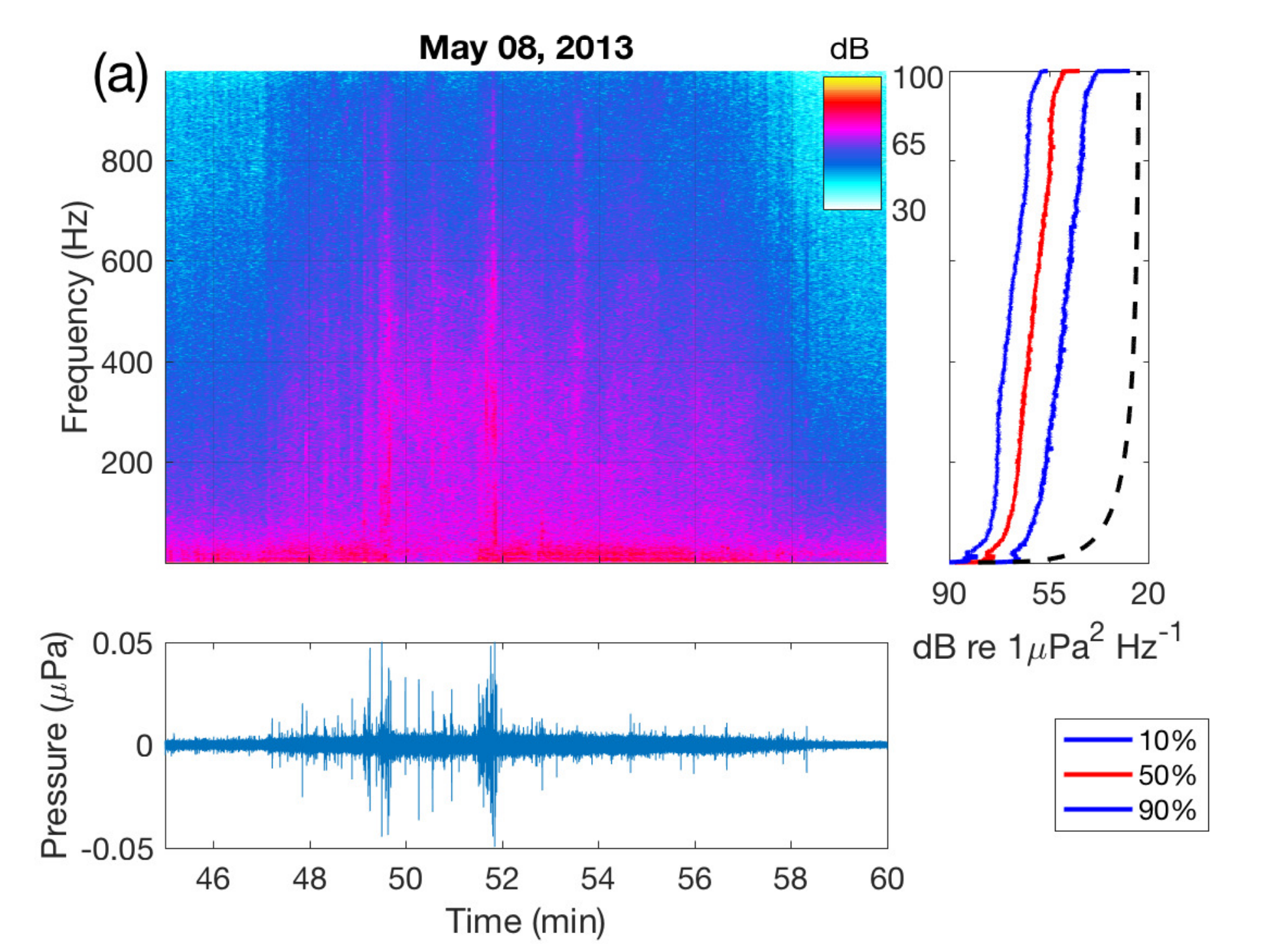}\label{fig: broadband_noise}}
\subfigure{
\includegraphics[width = 0.48\textwidth]{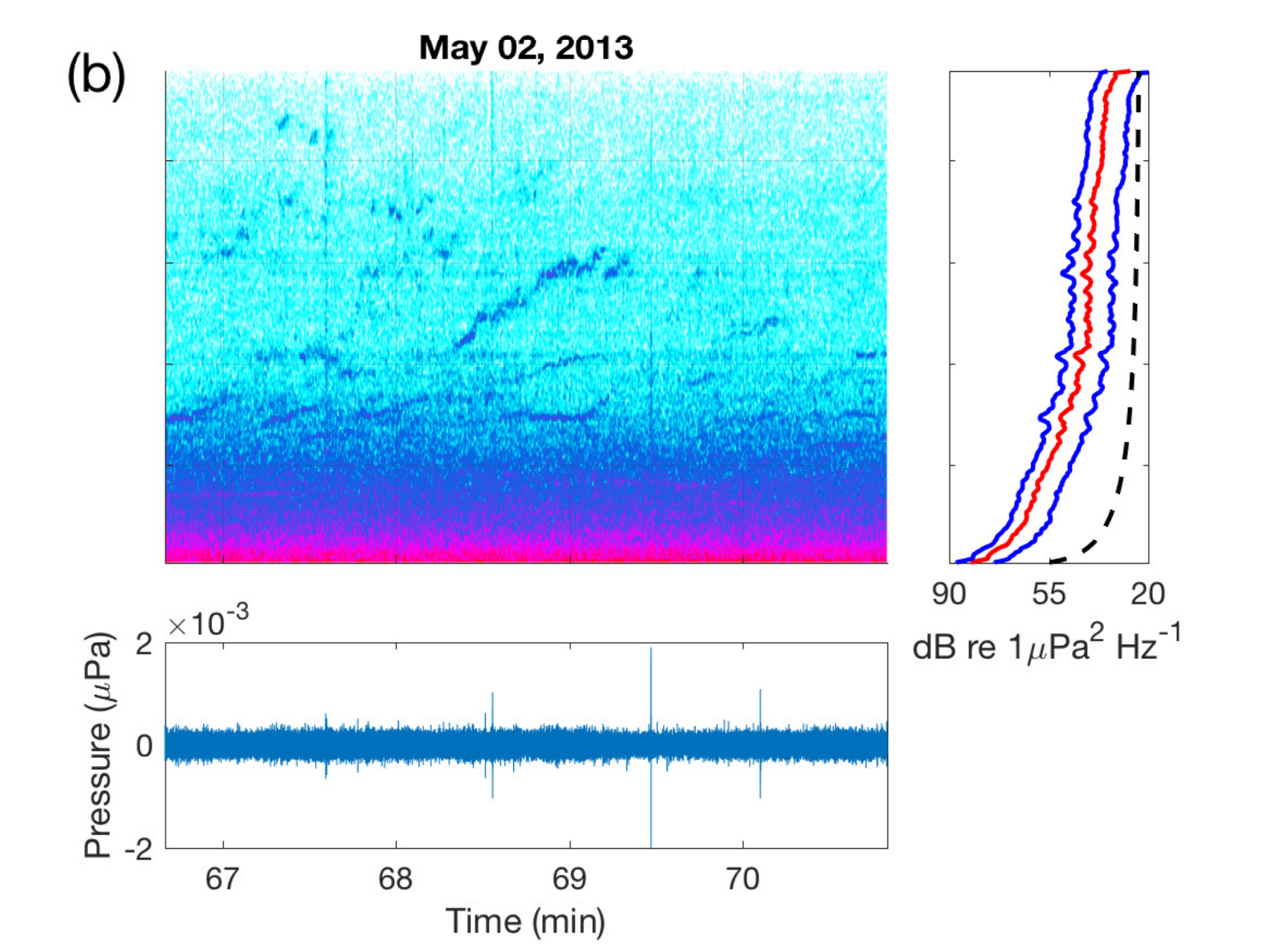}\label{fig: squeak1}}
\subfigure{
\includegraphics[width = 0.48\textwidth]{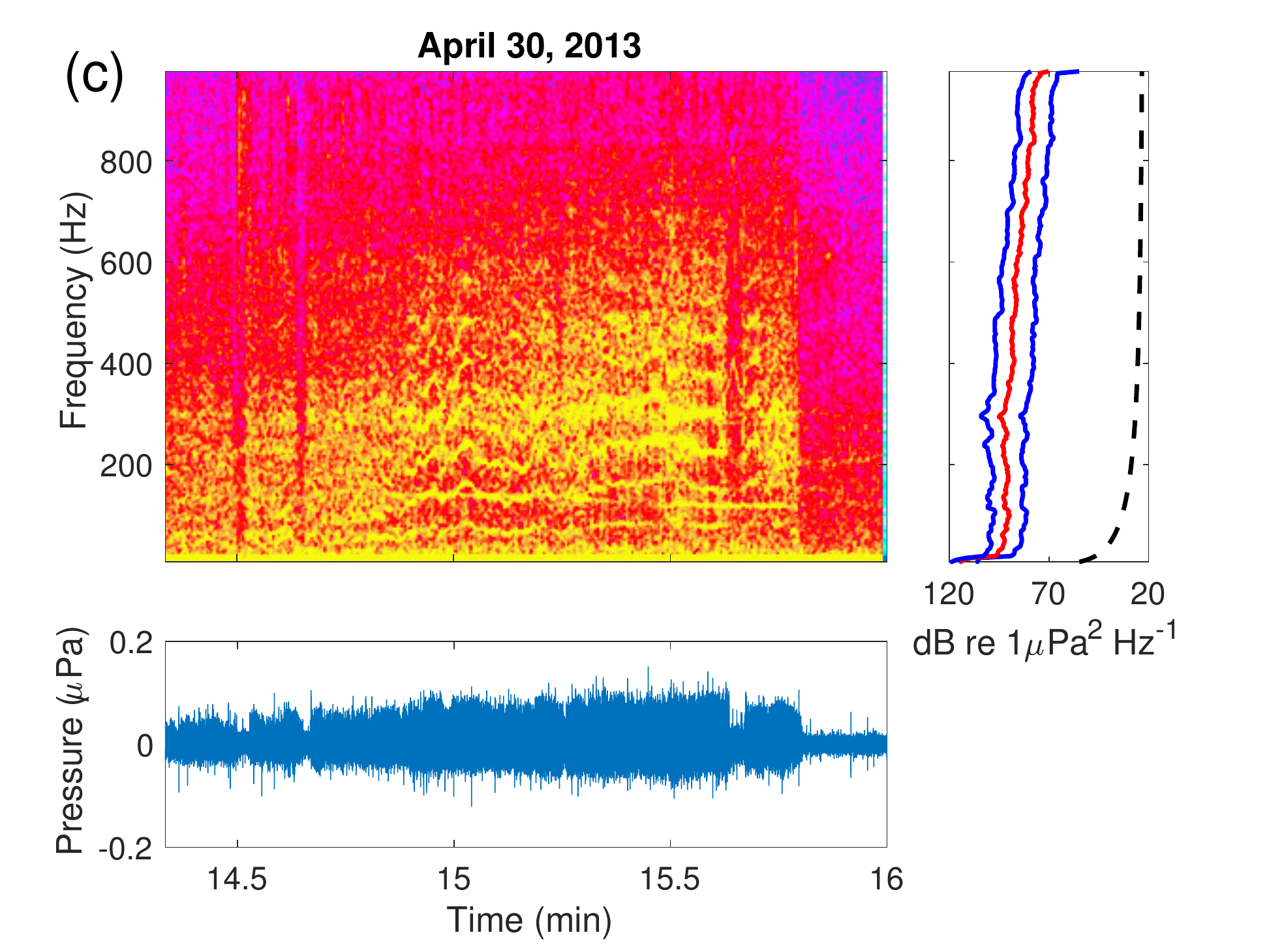}\label{fig: squeak2}}
\subfigure{
\includegraphics[width = 0.48\textwidth]{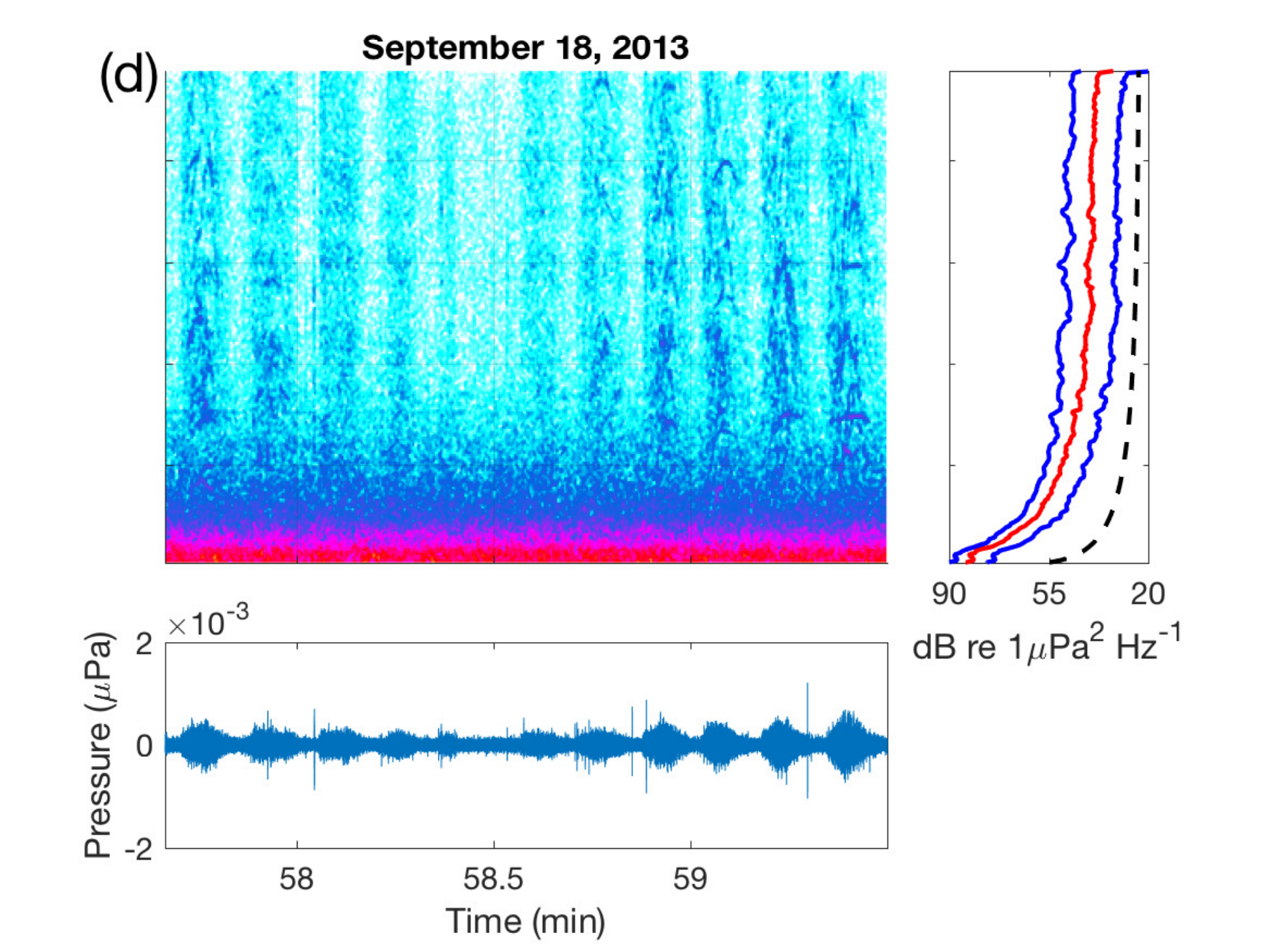}\label{fig: squeak3}}
\caption{(Color online) Spectrograms of ice noises including (a) a broadband event or events lasting up to 10 min, recorded on May 8, (b) non--constant tonals without harmonics lasting up to 1 min, recorded on May 2, (c) near-constant harmonic tonals lasting 2 min, recorded on April 30, and (d) non--constant, modulated harmonic tonals lasting for 5 s with a recurrent period on the order of ocean swell (9 s), recorded 230 km from the sea ice edge on September 18. The recording system noise is shown by the dashed black line.}
\label{fig: tonals}
\end{figure*}

Ice noises were observed to be either broadband or tonal in nature. Broadband noise generated by sea ice\cite{Kinda2015} appears as periods of elevated sound level, here ranging from 5--20 dB above the median level at 500 Hz (Fig.~\ref{fig: broadband_noise}) and lasting from 10--500 s. Broadband ice noise extended across the frequency band (Fig.~\ref{fig: broadband_noise}). Tonal ice noises are single--frequency or harmonic signatures modulated in time (Figs.~\ref{fig: squeak1}-\ref{fig: squeak3}).

Xie and Farmer\cite{XieFarmer91} demonstrated that constant--frequency ice tonals could be modeled as resonances in an infinitely long sea ice block of uniform height, density, and velocity generated by frictional shear stress on its edge. The non--constant tonals observed here may indicate anomalies in the local height or composition of the sea ice or a frictional stress that is velocity--dependent (Fig. \ref{fig: squeak1}). The slope and curvature of the tonals varies between hydrophone recordings (Fig. \ref{fig: squeak2}), indicating that significant changes in ice properties and dynamics may occur within the spatiotemporal span of 2--3 array drift days.  

Another interesting case are sets of modulated harmonics, ranging from 200--900 Hz, that are 8--10 dB louder than the background spectrum and last about 4 s, recurring with a period of about 9 s (Fig.~\ref{fig: squeak3}). These tonals may be due to ocean waves impinging on the sea ice edge, generating seismic or flexural waves that propagate within the sea ice if the product of the noise frequency and the sea ice thickness is less than about 300 Hz--m\cite{Stein88} and couple into the water column as periodically modulated harmonics. The observation of these tonals on the receiving array suggest that these effects can be seen at least as far as 230 km from the ice edge.

\subsection{Biological Sources\label{sec: Bio}}

\begin{figure}[]
\centering
\subfigure{
\includegraphics[width = 0.5\textwidth]{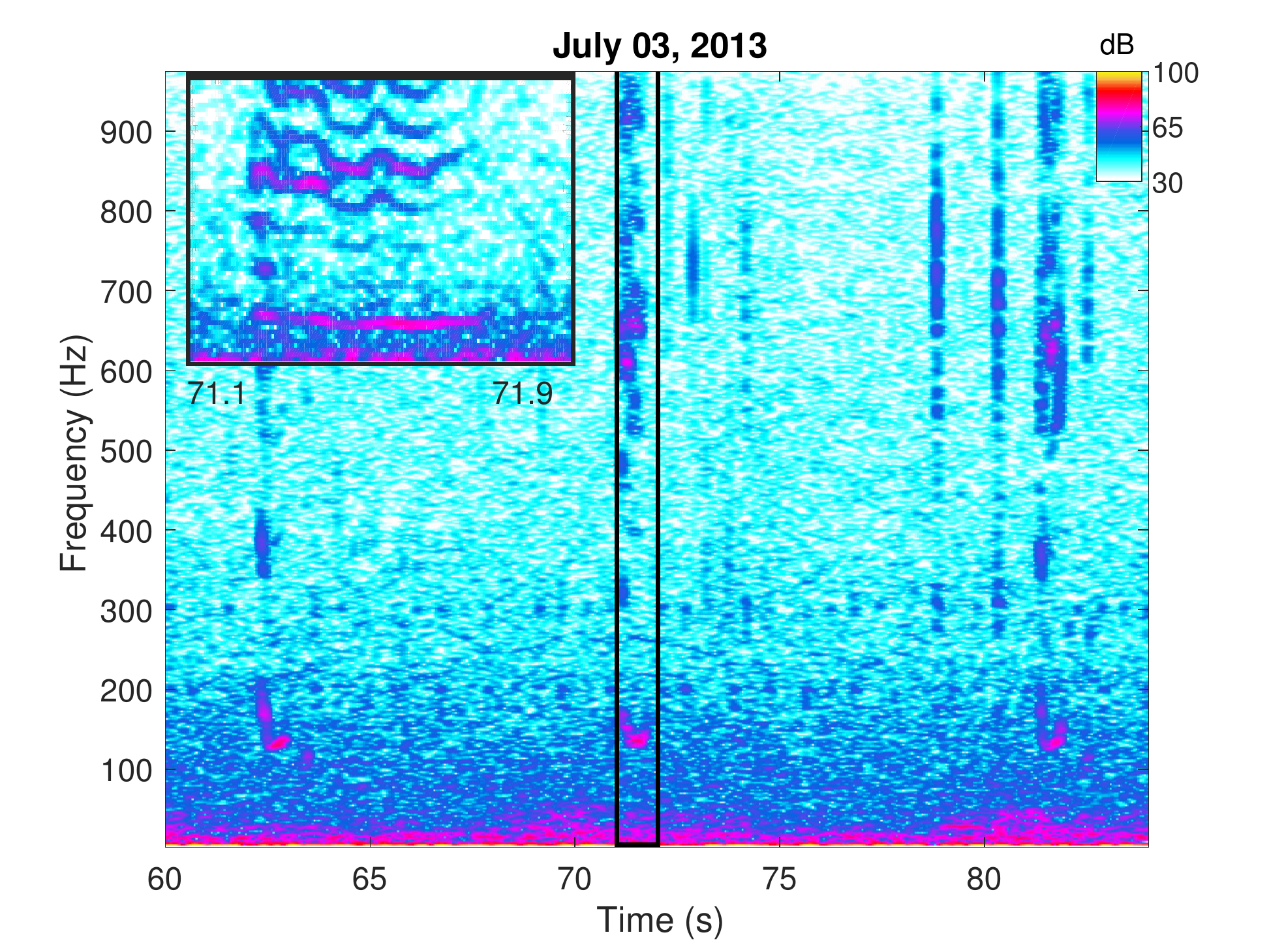}}
\caption{A series of calls from what is believed to be a Spitsbergen bowhead whale. The calling periodicity is about 10 s. These three calls were taken from a series lasting 55 s. The rectangle corresponds to the inset figure and shows a single call with harmonics from 150--976 Hz. The time axis in both figures is relative to 7 min in the recording on July 3.}\label{fig: bowhead}
\end{figure}

\begin{figure}[]
\centering
\includegraphics[width=0.5\textwidth]{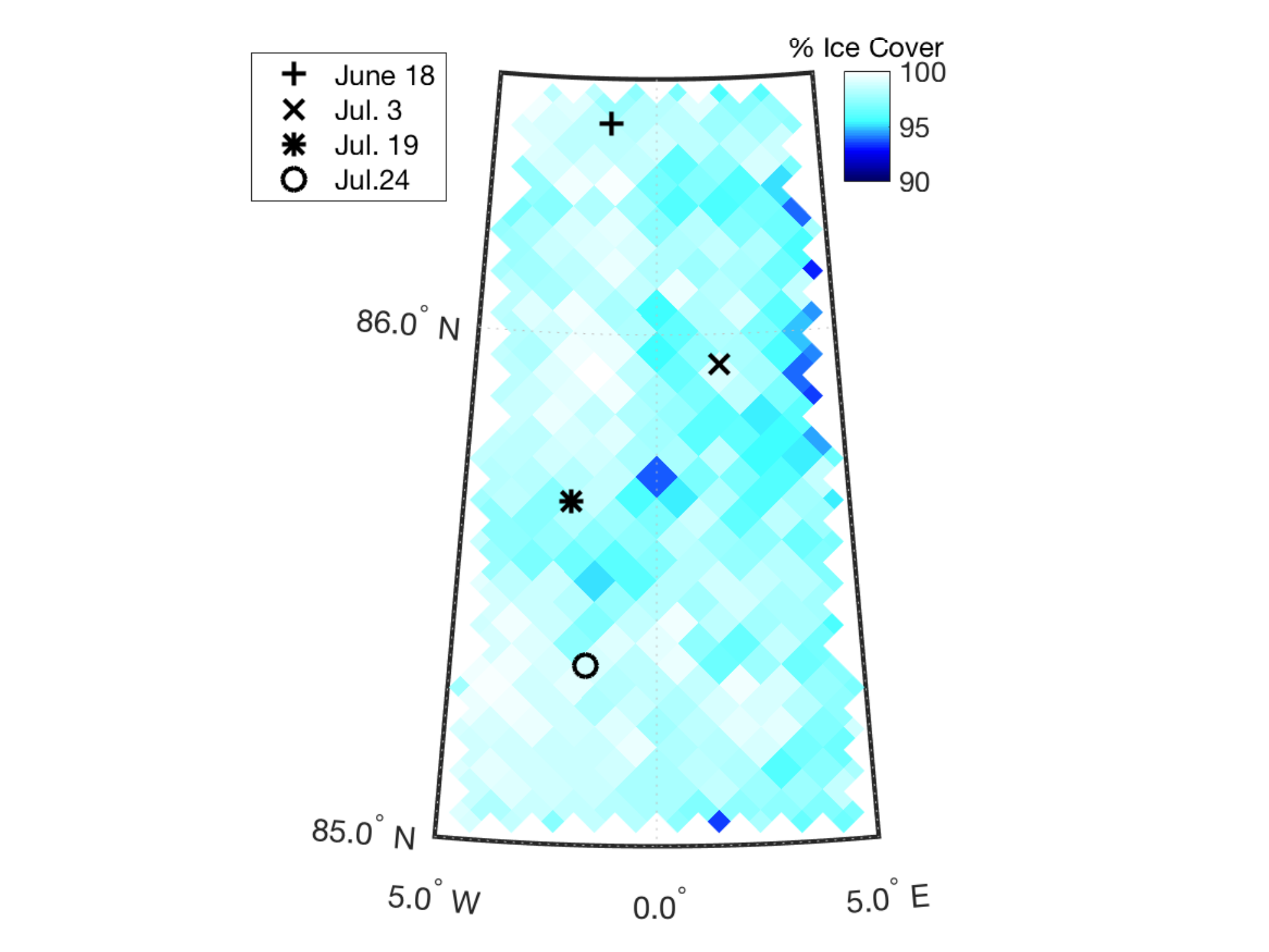}
\caption{AMSR2 satellite ice coverage averaged over the days when bowhead whale calls were recorded along with the location of the array on those days. Ice cover was close to 100\% at the array on these days.}\label{fig: ice_cover}
\end{figure}

Bowhead whale calls were observed during the summer 2013 array transit (Fig.~\ref{fig: bowhead}). The length of the call series lasted between 30 s and 7 min. The identification of the sound as a bowhead whale call was conducted by a manual analyst who led the team that identified thousands of bowhead whale calls in passive acoustic datasets recorded by instruments deployed during bowhead whale migrations along the North Slope of Alaska between 2008--2014.\cite{Blackwell2015}$^,$\cite{Thode2016} Calls were observed on June 18, July 3, 19, and 24. These calls were recorded when the array was northward of 85$\degree$N, at least 290 km north of other recordings in the region.\cite{Staffordetal2012}  Sea ice cover from AMSR2 satellite data\cite{AMSR2} was estimated to be higher than 90\% locally at the array for these days (Fig.~\ref{fig: ice_cover}).

Previous observations of bowhead whales have occurred southward of 82$\degree$30'N. Before the year 1818, the prolific species was fished in the region about 200 km west of Spitsbergen, between 76$\degree$N and 80$\degree$N. By 1818, this group had been depleted nearly to extinction.\cite{Staffordetal2012}  More recently, individuals or small groups have been acoustically detected as far north as 82$\degree$30'N.\cite{MooreReeves1993} Satellite--tagged whales in western Greenland spent most of their time in 90\% to 100\% ice cover far ($>$100km) inside the ice edge.\cite{Fergusonetal2010} A recent study of Spitsbergen bowhead whale calling near 78$\degree$50'N, 0$\degree$W recorded no calls between April 30 and September 1 in 2009.

Measurements of the relative timing of the whale call across the array aperture reveal that the animal was at least 50 km distant. However, placing an upper bound on the range is difficult. Using received levels to estimate source range is imprecise for two reasons:  the bowhead whales are capable of calling across a broad spread of source levels\cite{Thode2016}, and uncertainties arise rise when modeling transmission loss due to scattering of signals from ice.  Using timing measurements of signal arrivals across the array for localization is feasible, but requires that the vertical array tilt and sound speed profile be modeled or inverted correctly, a topic beyond the scope of the present paper.

\subsection{Seismic Survey Signals\label{sec: Airguns}}

Broadband pressure pulses generated by airguns are used to image the geological structure beneath the seafloor during seismic surveys. At long distances, frequencies higher than about 100 Hz are attenuated. The resulting pulses are observed on hydrophone receivers at frequencies below 50 Hz. Distant noise from seismic surveys can be observed almost daily in the Fram Strait during summer months. E.g. in a previous dataset in the Fram Strait, airgun surveys were observed on 90--95\% of days between July and September 2009.\cite{Moore2012}

\begin{figure}[]
\centering
\subfigure{
\includegraphics[width = \linewidth]{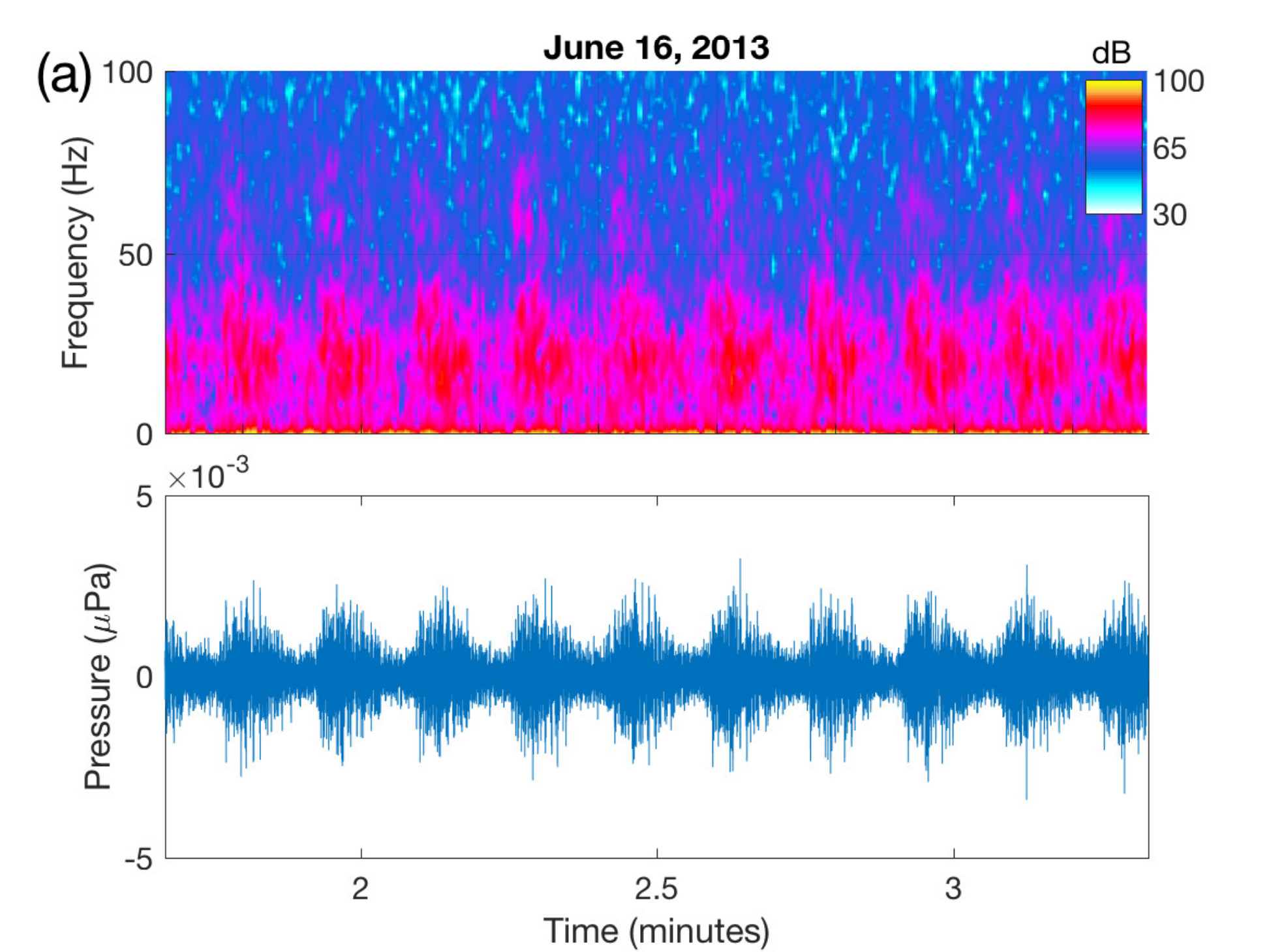}\label{fig: airgun}}
\subfigure{
\includegraphics[width = \linewidth]{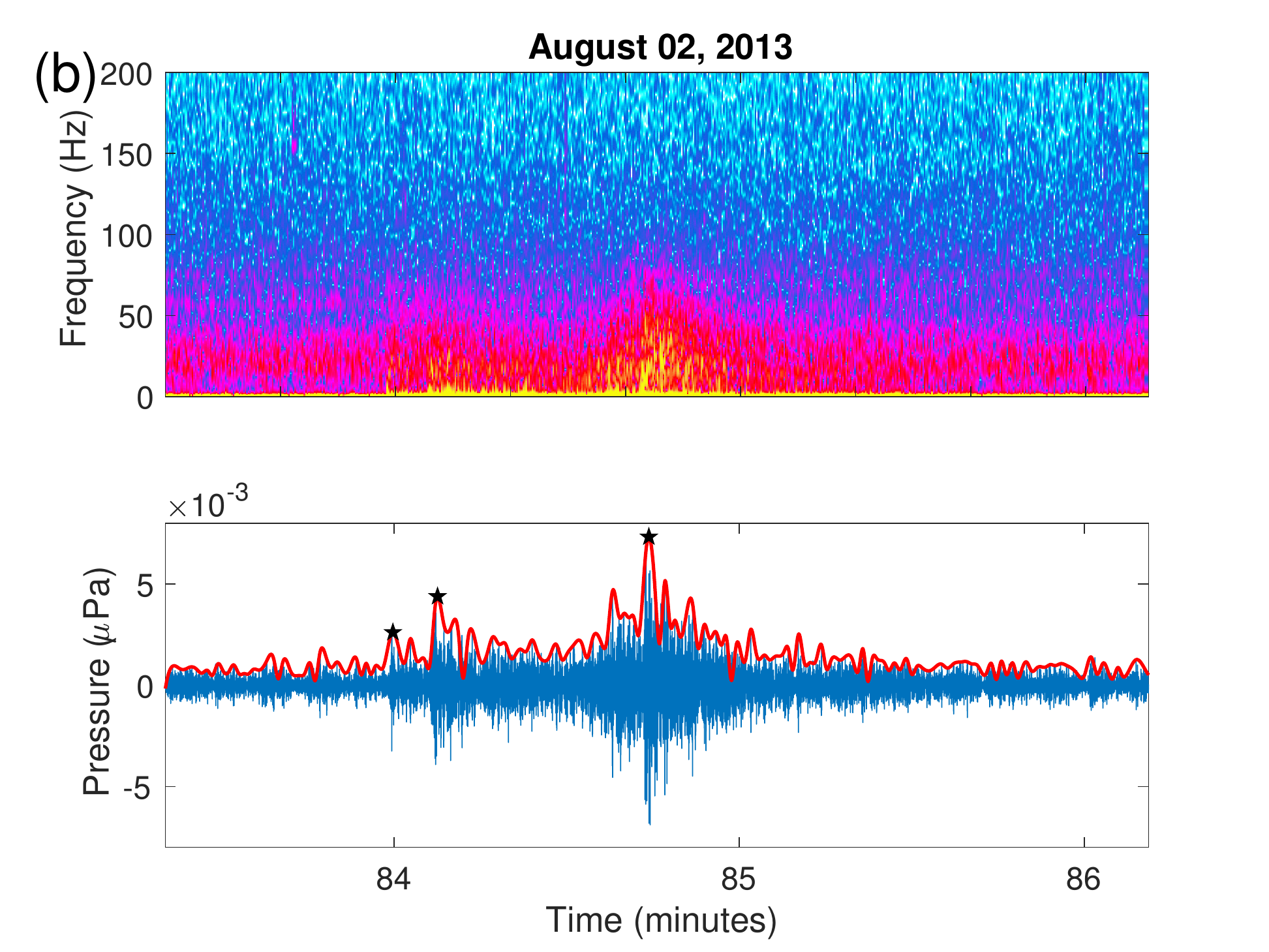}\label{fig: earthquake}}
\caption{(Color online) Spectrograms and time series of (a) low--frequency pulses generated by a distant airgun survey recorded on June 16 and (b)  an earthquake recorded on August 2, where the wave arrival delays are used to estimate source range.}
\end{figure}

In this dataset, airgun pulses were observed between May 7 and Sep. 19 and were present on 11 of the 19 recording days  (Fig. \ref{fig: airgun}), with nearly continuous pulses detected during the 108 min recording period whenever observed. Location, type and date of surveys in Norwegian territory were obtained from the Norwegian Petroleum Directorate. According to these data, the array was 1800--3500 km distant from seismic surveys at the start in April and 1000--3000 km distant at the end in September. Seismic surveys conducted in the Canadian Arctic during summer 2013 may have been detected, but survey details were not publicly available.

Transmission loss estimates across the MIZ near the Fram Strait, extending as far as 150 km into the ice, have demonstrated that the under--ice transmission loss is smaller than previously proposed at low frequencies.\cite{TollefsenSagen}$^,$\cite{LePageSchmidt94} The observations here also suggest that the change in transmission loss far into the compact ice is small, but uncertainties in source spectrum and distance make quantitative transmission loss estimates unreliable.

\subsection{Arctic Basin Earthquakes}

Hydrophone arrays are valuable earthquake monitoring tools. The acoustic $T$--phase pressure wave (see Fig.~\ref{fig: earthquake}) is coupled into the water column at a seamount or down--sloping bathymetric feature near the earthquake. The versatility of hydrophone arrays enables them to be deployed in difficult areas such as the active Gakkel Ridge in the ice--covered Arctic, where ocean bottom seismometers are challenging to deploy.\cite{SohnHildebrand2001}

Time difference of arrival between the $T$, $P$, and $S$ arrivals can be used on a hydrophone array to estimate the earthquake distance:
\begin{equation}\label{eq: 1}
 R = \frac{\Delta \tau}{\Big(\frac{1}{v_{T}} - \frac{1}{v_{P}}\Big)} 
\end{equation}

 where $R$ is the range to the earthquake, $\Delta \tau$ is the arrival time difference, $v_T$ is the group velocity of the $T$--phase, and $v_P$ is the group velocity of the $P$ wave (or $S$ wave).

Three $T$--phase arrivals were observed during the array transit along with occasional $P$ and $S$ wave arrivals (Fig. \ref{fig: earthquake}). Overall, three $T$--phase events were identified in the data, each lasting 1 min. The arrivals in Fig.~\ref{fig: earthquake} are applied to the time difference method in Eq. \ref{eq: 1} with $v_T$ from the CTD measurement (1.44 km/s) at deployment and $v_P$, $v_S$ (6.1 and 3.1 km/s) estimated from the IASPEI seismic catalogue and adjusted to achieve agreement between estimates. Although the travel time of the $T$--phase may be biased depending on where it couples into the water column, the estimated earthquake distances of 90 km for the $P$--$T$ difference and 100 km for $S$--$T$ difference agree well here.

The earthquake distance estimate indicates that the event originated at the Gakkel Ridge. The earthquake was not registered in the Global Seismic Network catalogue which only records events with $m_b > 4$. The detection of $T$, $P$, and $S$ arrivals on a single hydrophone for an unregistered earthquake demonstrates the potential for underwater acoustic monitoring of low magnitude seismic activity near the Gakkel Ridge.

\section{Arctic ambient noise levels}\label{sec: results}

\subsection{Eastern Arctic Ambient Noise, Summer 2013}\label{subsec: NoiseProcessing}

Statistical analyses were conducted for three and four day periods across the array drift path: May 1, 2, 7; May 8, 9, 12, 14; June 16, 18, July 3, 14; July 19, 24, August 2; and September 10, 18, 19. April 30 and September 20 contain anomalous ice and ship noise events and are excluded from the statistical analyses.

\begin{figure}[]
\centering
\includegraphics[width = \linewidth]{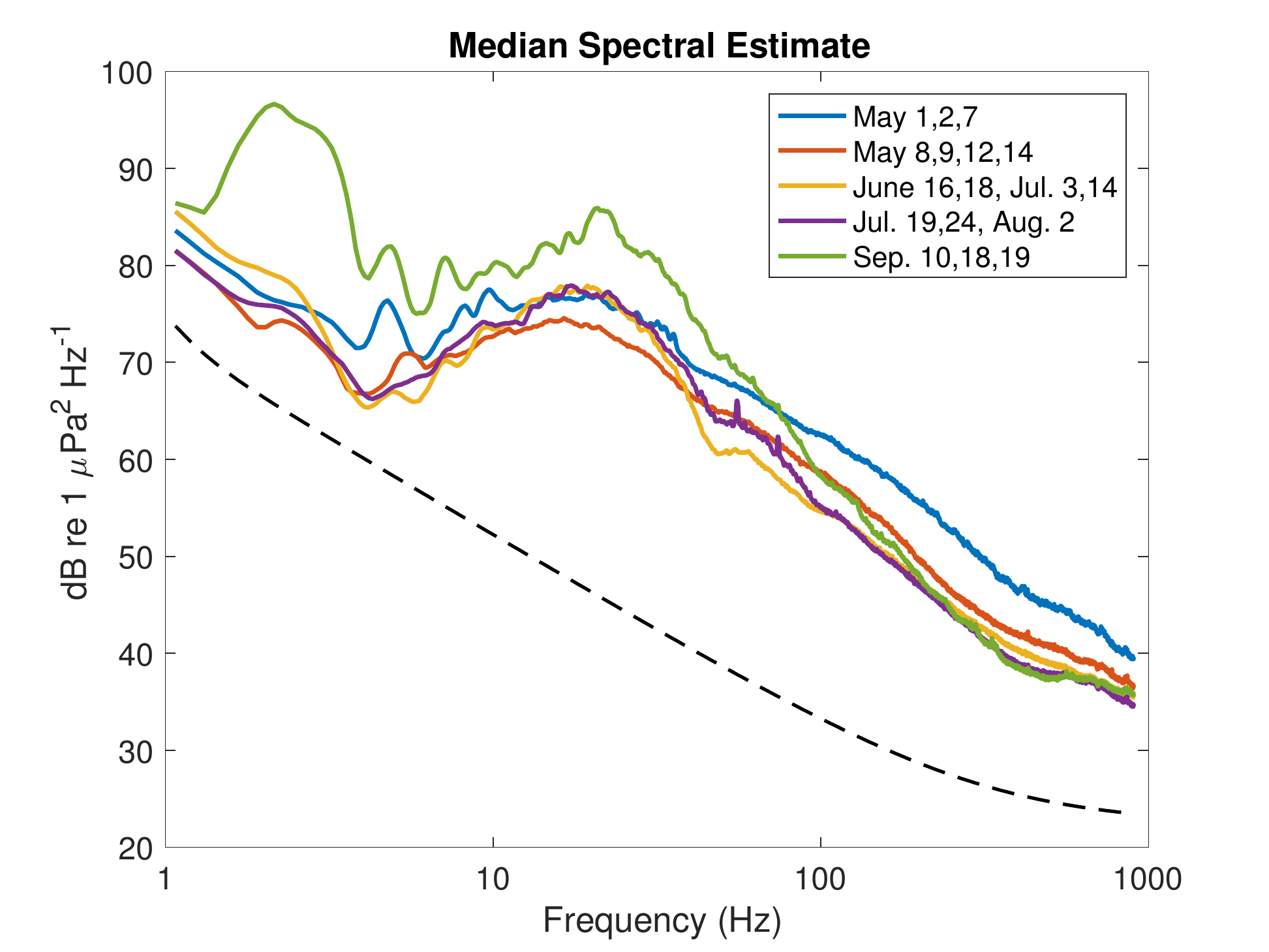}
\caption{(Color online) Median power spectral estimates for three and four day periods in summer 2013. The recording system noise is shown by the dashed black line.}\label{fig: dailymonthly}
\end{figure}

The median power spectra show characteristics of Arctic ambient noise and its sources (Fig.~\ref{fig: dailymonthly}). The broad peak at 15--20 Hz is attributed to the ice--scattered propagation characteristics of distant sources,\cite{MakrisDyer86} as higher frequencies are more attenuated and lower frequencies have bottom interacting modes (see Sec.~\ref{sec: prop}). Seismic airgun surveys increase the median power at frequencies between about 10 Hz and 100 Hz (Fig.~\ref{fig: dailymonthly}) due to the dispersive quality of the pulse arrivals. Observations of the spectrogram estimates confirm that the increase in low frequency power for September results from an increase in the received levels of airgun pulses. Likewise, decreased low frequency power in the May 8, 9, 12, 14 period results from lulls in the presence of airgun noise. Transient ice noises result in elevated power levels for frequencies above 100 Hz (Fig.~\ref{fig: dailymonthly}). Transient ice noises were observed in the spectrograms estimates most frequently and at the highest received levels during May 1, 2, 7 and May 8, 9, 12, 14.

\begin{figure}[]
\centering
\includegraphics[width = \linewidth]{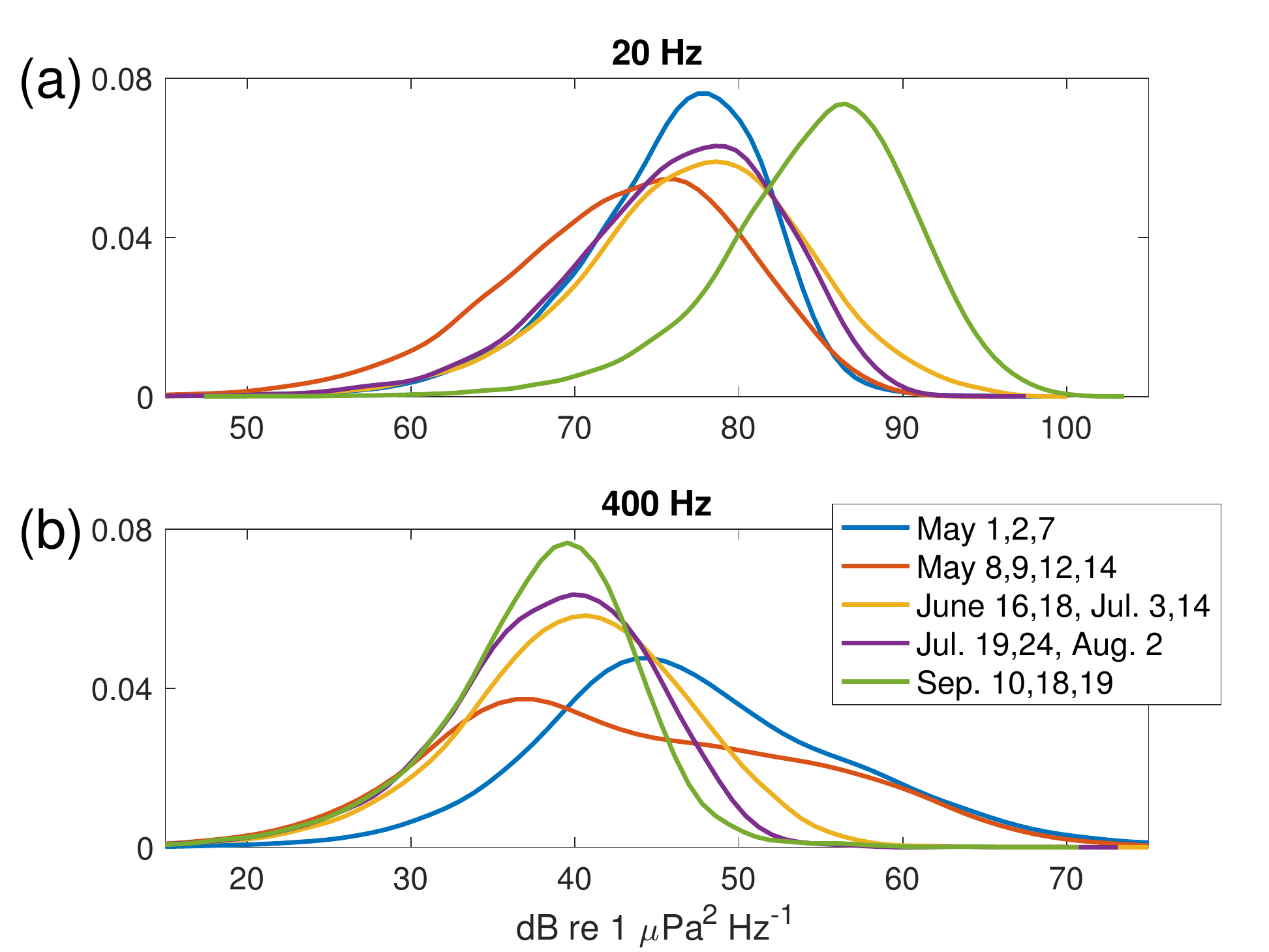}
\caption{(Color online) Empirical probability density functions (PDFs) estimated for the three and four day periods in summer 2013 at (a) 20 Hz and (b) 400 Hz. The 20 Hz estimate is predominantly effected by presence and strength of airgun pulse noise while the 400 Hz estimate corresponds to transient ice noises.}\label{fig: PDFmonthly}
\end{figure}

The empirical probability density functions (PDFs) were estimated at 20 Hz and 400 Hz  (Figs.~\ref{fig: PDFmonthly}(a) and Fig.~\ref{fig: PDFmonthly}(b), see Sec.~\ref{subsec: acoustics} for details). At 20 Hz, the variation in the median power level corresponds to changes in the received level of seismic airgun noise. May 8, 9, 12, 14 also exhibits a broader distribution as a result of the lull in airgun noise during this period (Fig.~\ref{fig: PDFmonthly}(a)). At 400 Hz, the distributions for May 1, 2, 7 and May 8, 9, 12, 14 are highly non--Gaussian as a result of numerous, loud transient ice noise events (Fig.~\ref{fig: PDFmonthly}(b)). During the remaining periods, ice noises were received at lower and more consistent power levels, resulting in more peaked distributions.

\begin{figure}[]
\centering
\includegraphics[width =  \linewidth]{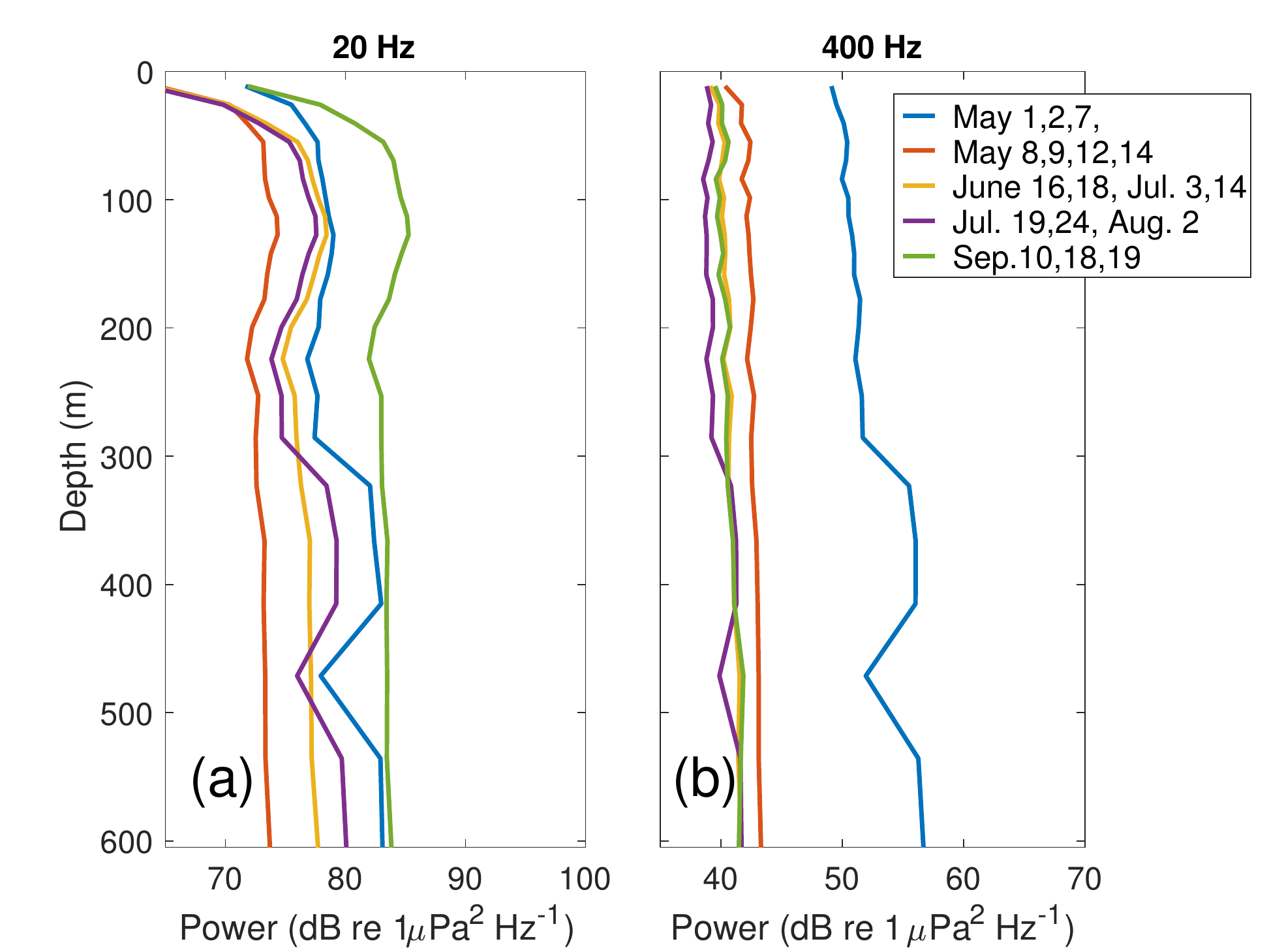}
\caption{(Color online) Depth dependence of median spectral power for three and four day periods in summer 2013 for (a) 20 Hz and (b) 400 Hz indicates that the effect of noise sources is consistent with depth.}
\label{fig: depth_lof}
\end{figure}

Median estimates for all hydrophones on the array show that the effect of noise sources is consistent with depth. At 20 Hz (Fig.~\ref{fig: depth_lof}(a)) the median estimates in depth reflect the shapes of the first and second mode (see Sec.~\ref{sec: prop}, Fig.~\ref{fig: Modes}). The 400 Hz median estimates are nearly constant in depth (Fig.~\ref{fig: depth_lof}(b)), with the May 1, 2, 7 and May 8, 9, 12, 14 estimates at elevated power levels. Increased power levels below 300 m at both frequencies (Fig.~\ref{fig: depth_lof}) may be evidence that the effort to eliminate flow--related noise artifacts was not completely successful for all hydrophones and periods.

\subsection{Comparison of Arctic Ambient Noise}

\begin{figure}[]
\centering
\includegraphics[width =  \linewidth]{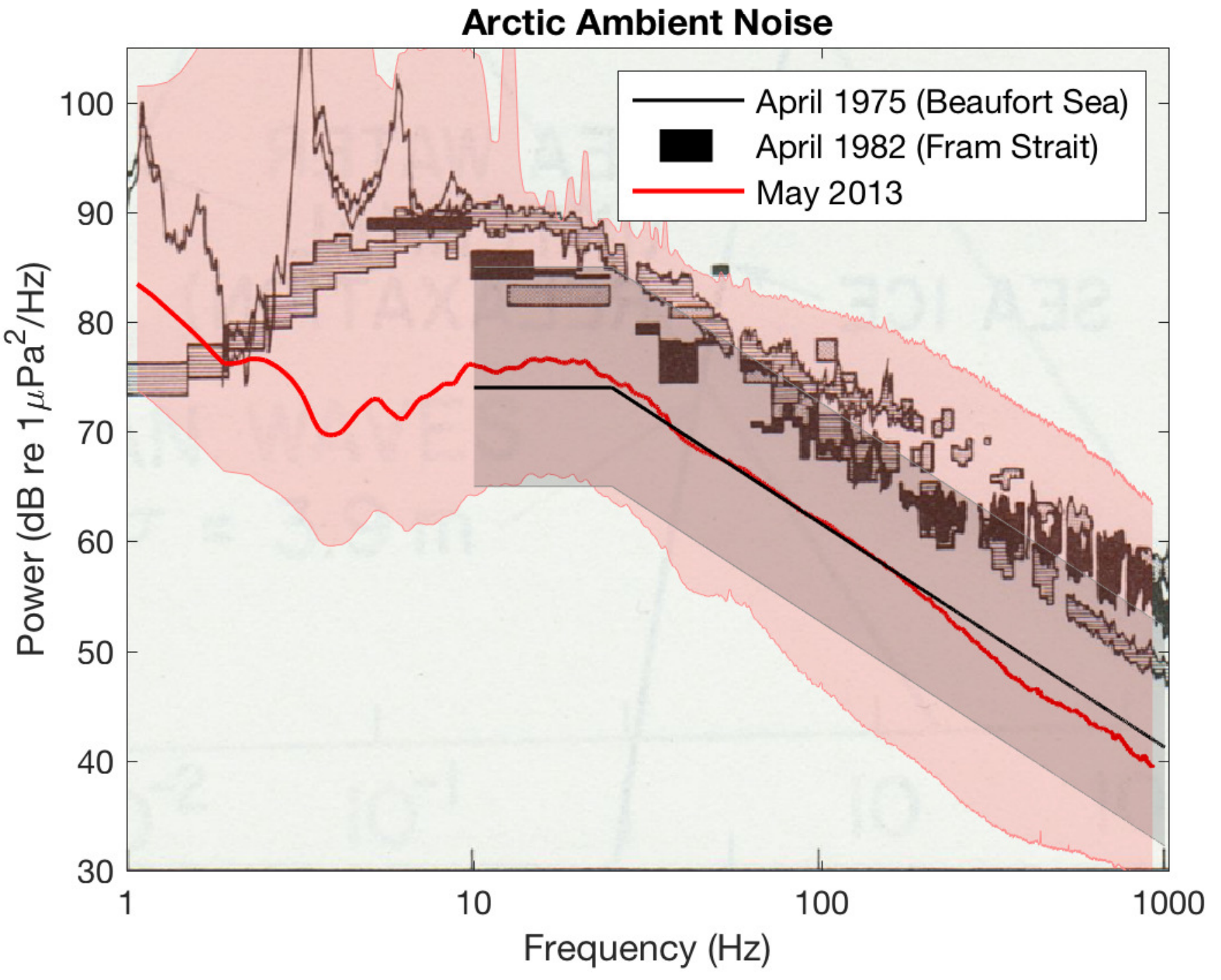}\label{fig: THAAW_FRAMIV}
\caption{(Color online) Median spectral estimate for May 2013 at 84.5 m depth,  April 1982 (FRAM IV, Beaufort Sea)\cite{MakrisDyer86} at 99 m depth, and April 1975 Polar Research Laboratory (Beaufort Sea, depth not published)\cite{BuckWilson86} (see Table \ref{tab: SpectralLevels}). The 10\% and 90\% spectral levels for May 2013 are shaded on either side of the median estimate; 5\% and 95\% are given in a smaller shaded region for April 1975. The FRAM IV estimate is a composite of measurements taken at different times and averaged over various numbers of samples and frequency bandwidths.\cite{MakrisDyer86}}\label{fig: comparisons}
\end{figure}

\begin{figure}[]
\includegraphics[width = \linewidth]{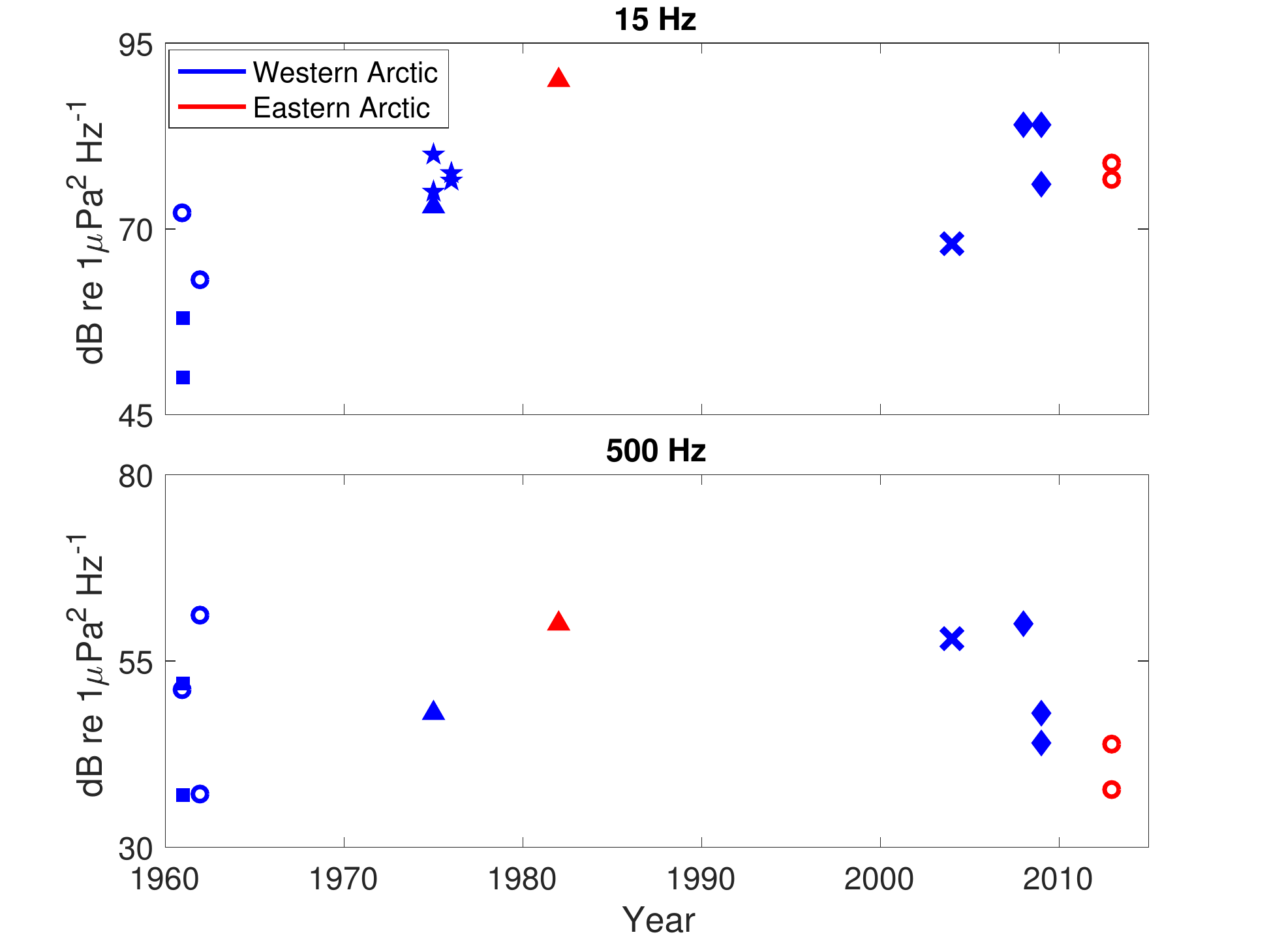}
\caption{(Color online) Scatter plot of median ambient noise level results for 15 Hz and 500 Hz from various studies in both the eastern and western Arctic (see Table \ref{tab: SpectralLevels}).}\label{fig: Results}
\end{figure}

\begin{table*}[]
\centering
\caption{\label{table: comparison_chart} Ambient noise noise level estimates in the Arctic Ocean.}
             \begin{tabular}{ccccccccc}
             \hline\hline
              Location (lat, lon) & Experiment & Dates  & 15 Hz & 50 Hz & 100 Hz & 500 Hz & 1 kHz \\[0.5ex] 
             \hline
	 	 86\degree N 56.9\degree W --   & \includegraphics[width = 0.15in]{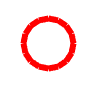} May--June 2013 & 05/2013 --  & 76.5 & 66 & 60.2 & 43.7 & - \\
		  89\degree N  1\degree E  &   & 06/2013   &    &    &   &     &     \\
	 	 86\degree N 1.3\degree E --  &  \includegraphics[width = 0.15in]{red_dot} July--Sep. 2013 & 07/2013 --  & 78.7 & 64.9 & 55.6 & 37.6 &  -  \\
		 83.8\degree N 4.5\degree E &      &   09/2013  &    &    &    & &   \\
		
	 	 83\degree N 20\degree E & \includegraphics[width = 0.15in]{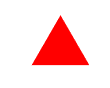} FRAM IV\cite{MakrisDyer86} & 04/1982 & 90 & 79.5 & 73 & 60  & 53  \\\hline
		 
		 		 82\degree N 168\degree E   &  \includegraphics[width = 0.15in]{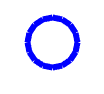} Mellen, Marsh 1985\cite{MellenMarsh85} & 09--10/1961  & 72 & 70 & 61 & 51 & 40 \\
		  75\degree N 168\degree W &  & 05--09/1962   & 63 & 64 & 49 & 37 & 32  \\
		  &  &  & - & 75 & 72 & 61 & 52 \\
		 	78.5\degree N 105.25\degree W  & \includegraphics[width = 0.1in]{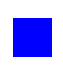} Ice Pack I \cite{MilneGanton64}  &  27/04/1961 & 50 & 42 & 38 & 37 & 20  \\
		&   & 28/04/1961  & 58 & 52 & 51 & 52 & 51 \\
		 74.5\degree N 115.1\degree W  &  \includegraphics[width = 0.1in]{blue_square} Ice Pack II \cite{MilneGanton64} &  9/2--3/1961 & - & 57 & 56 & 52 & 43 \\
		 		 
		  Beaufort Sea &  \includegraphics[width = 0.15in]{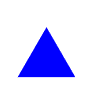} PRL\cite{BBuck} & April 1975  & 73 & 68 & 62 & 48 & 43  \\
		 &&& (10 Hz) & (32 Hz) &&&\\
		 $\sim$72\degree N 142\degree W &\includegraphics[width = 0.15in]{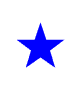} AIDJEX\cite{LewisDenner87seasonal} & 08/1975 & 65--85 & 65--75 & - & - & 38--55 \\
		  & & 11/1975 & 70--90 & 65--88 & - & -  & 40--70 \\
		  &  & 02/1976 & 65--90 & 60--90 & - & - & 35--70 \\
		  & & 05/1976 & 65--88 & 60--90 & - & - & 37--68 \\
		 
		 		 71\degree N 126.07\degree W  &  \includegraphics[width = 0.1in]{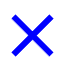} Kinda et al. 2013\cite{Kinda2013} & 11/2004 --& 68 & 69 & 66 & 58 & 54  \\
		 & &  06/2005  &  & & & & \\
		 
		  72.46\degree N 157.4\degree W &  \includegraphics[width = 0.15in]{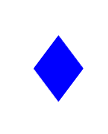}  Roth et al. 2011\cite{Roth} &  09/2008 & 84 & 80 & 74 & 60 & 56  \\
		   & & 03/2009 & 84 & 70 & 62 & 48 & 48  \\
		 & & 05/2009 & 76 & 61 & 56 & 44 & 44  \\
		
		 \hline\hline
		\end{tabular}
	\label{tab: SpectralLevels}
\end{table*}

The median spectral power across the period including April 30, May 1, 2, 7--9, 12, and 14 at 84.5 m depth are compared with historical estimates from both western and eastern Arctic stations in Fig.~\ref{fig: comparisons}.
The estimated median spectral power for May 2013 was below, but similarly structured to, a composite spectral estimate from April 1982 (Fig.~\ref{fig: comparisons}).\cite{MakrisDyer86} The peak at 15 Hz appears less prominent at lower frequencies in 2013 than in 1982. In comparison, a spectral estimate recorded in the Beaufort Sea in April 1975 shows comparable ambient noise levels and structure to 2013 but does not extend to lower frequencies (Fig. \ref{fig: comparisons}).\cite{BuckWilson86} The differences in these spectra may be caused by environmental factors or by experimental factors, including recording length and post--processing methods, which were not published alongside the 1982 results.

Fig. \ref{fig: Results} demonstrates the wide variability in Arctic ambient noise estimates across frequency, year, and study. This variability arises from a complex relationship between the Arctic ambient noise and both environmental and anthropogenic factors, such as sea ice percent cover, sea ice age/thickness, barometric conditions and wind patterns, local subsurface currents, seismic survey activity, and marine biologic activity. The studies shown indicate that, without correction for environmental factors, there is not a significant trend in the Arctic ambient noise power levels between 1960 and 2013, but that frequency--dependent ambient noise levels are within a 30--40 dB range for both regions of the ice covered Arctic.

\section{Conclusions}
Between April and September 2013, a twenty--two element vertical hydrophone array recorded the eastern Arctic ambient noise for 108 min/day while drifting between 89\degree N, 62\degree W and Svalbard.

These data were processed into spectrograms and a number of noise sources were observed, including ice noise, bowhead whale calling, airgun survey pulses, and earthquake $T$--phases. The bowhead whale calls were received between 86 and 87$\degree$ N in June and July.

The data were also processed into three and four day median spectral estimates. The spectral estimates and corresponding PDFs demonstrate the variation in the occurrence and received level of seismic airgun survey pulses at low frequencies and ice transients at high frequencies.

The median spectral estimate for May 2013 was compared to historical power spectral estimates, one  recorded in a nearby region in April 1982 \cite{MakrisDyer86} and another from an ice--covered region in the Beaufort Sea in April 1975.\cite{BuckWilson86} The May 2013 estimate is below the 1982 estimate but close to the 1975 estimate, indicating that local ice source effects may be as significant as regional effects in determining ambient noise levels in the Arctic. A multi--decadal summary of Arctic ambient noise studies displays a lack of change in power levels with time and further demonstrates the variability in Arctic ambient noise level estimates resulting from local experimental variations.

\begin{acknowledgments}
We would like to thank John Colosi for providing the MicroCAT data, John Kemp and the WHOI Mooring Operations and Engineering Group for their assistance, and Hanne Sagen and the Norwegian Coast Guard for assistance in recovering the array mooring. This work is supported by the Office of Naval Research under Award Numbers N00014-13-1-0632 and N00014-12-1-0226. Any opinions, findings, and conclusions or recommendations expressed in this publication are those of the authors and do not necessarily reflect the views of the Office of Naval Research.
\end{acknowledgments}

\bibliographystyle{unsrt}

\end{document}